\documentstyle[12pt,epsf]{article}
\begin{document}

\catcode`@=11
\long\def\@caption#1[#2]#3{\par\addcontentsline{\csname
  ext@#1\endcsname}{#1}{\protect\numberline{\csname
  the#1\endcsname}{\ignorespaces #2}}\begingroup
    \small
    \@parboxrestore
    \@makecaption{\csname fnum@#1\endcsname}{\ignorespaces #3}\par
  \endgroup}
\catcode`@=12
\newcommand{\newc}{\newcommand}
\newc{\gsim}{\lower.7ex\hbox{$\;\stackrel{\textstyle>}{\sim}\;$}}
\newc{\lsim}{\lower.7ex\hbox{$\;\stackrel{\textstyle<}{\sim}\;$}}
\newc{\gev}{\,{\rm GeV}}
\newc{\mev}{\,{\rm MeV}}
\newc{\ev}{\,{\rm eV}}
\newc{\kev}{\,{\rm keV}}
\newc{\tev}{\,{\rm TeV}}
\def\tr{\mathop{\rm tr}}
\def\Tr{\mathop{\rm Tr}}
\def\Im{\mathop{\rm Im}}
\def\Re{\mathop{\rm Re}}
\def\bR{\mathop{\bf R}}
\def\bC{\mathop{\bf C}}
\def\lie{\mathop{\hbox{\it\$}}} 
\newc{\sw}{s_W}
\newc{\cw}{c_W}
\newc{\swsq}{s^2_W}
\newc{\swsqb}{s^2_W}
\newc{\cwsq}{c^2_W}
\newc{\cwsqb}{c^2_W}
\newc{\Qeff}{Q_{\rm eff}}
\newc{\fpf}{{\bf\bar5}+{\bf5}}
\newc{\tpt}{{\bf\overline{10}}+{\bf10}}
%
%
\def\NPB#1#2#3{Nucl. Phys. {\bf B#1} (19#2) #3}
\def\PLB#1#2#3{Phys. Lett. {\bf B#1} (19#2) #3}
\def\PLBold#1#2#3{Phys. Lett. {\bf#1B} (19#2) #3}
\def\PRD#1#2#3{Phys. Rev. {\bf D#1} (19#2) #3}
\def\PRL#1#2#3{Phys. Rev. Lett. {\bf#1} (19#2) #3}
\def\PRT#1#2#3{Phys. Rep. {\bf#1} (19#2) #3}
\def\ARAA#1#2#3{Ann. Rev. Astron. Astrophys. {\bf#1} (19#2) #3}
\def\ARNP#1#2#3{Ann. Rev. Nucl. Part. Sci. {\bf#1} (19#2) #3}
\def\MPL#1#2#3{Mod. Phys. Lett. {\bf #1} (19#2) #3}
\def\ZPC#1#2#3{Zeit. f\"ur Physik {\bf C#1} (19#2) #3}
\def\APJ#1#2#3{Ap. J. {\bf #1} (19#2) #3}
\def\AP#1#2#3{{Ann. Phys. } {\bf #1} (19#2) #3}
\def\RMP#1#2#3{{Rev. Mod. Phys. } {\bf #1} (19#2) #3}
\def\CMP#1#2#3{{Comm. Math. Phys. } {\bf #1} (19#2) #3}
\relax
%
%
%
\def\beq{\begin{equation}}
\def\eeq{\end{equation}}
\def\bea{\begin{eqnarray}}
\def\eea{\end{eqnarray}}
%
%
%
\def\boxeqn#1{\vcenter{\vbox{\hrule\hbox{\vrule\kern3pt\vbox{\kern3pt
\hbox{${\displaystyle #1}$}\kern3pt}\kern3pt\vrule}\hrule}}}
%
%
\def\mbox#1#2{\vcenter{\hrule \hbox{\vrule height#2in
\kern#1in \vrule} \hrule}}
\def\half{{\textstyle{1\over2}}} 
%
%
%
%
\newc{\ie}{{\it i.e.}}          \newc{\etal}{{\it et al.}}
\newc{\eg}{{\it e.g.}}          \newc{\etc}{{\it etc.}}
\newc{\cf}{{\it c.f.}}
%
%
%
%
\def\CAG{{\cal A/\cal G}} 
\def\CA{{\cal A}} \def\CB{{\cal B}} \def\CC{{\cal C}} \def\CD{{\cal D}}
\def\CE{{\cal E}} \def\CF{{\cal F}} \def\CG{{\cal G}} \def\CH{{\cal H}}
\def\CI{{\cal I}} \def\CJ{{\cal J}} \def\CK{{\cal K}} \def\CL{{\cal L}}
\def\CM{{\cal M}} \def\CN{{\cal N}} \def\CO{{\cal O}} \def\CP{{\cal P}}
\def\CQ{{\cal Q}} \def\CR{{\cal R}} \def\CS{{\cal S}} \def\CT{{\cal T}}
\def\CU{{\cal U}} \def\CV{{\cal V}} \def\CW{{\cal W}} \def\CX{{\cal X}}
\def\CY{{\cal Y}} \def\CZ{{\cal Z}}
%
%
%
%
%
\def\grad#1{\,\nabla\!_{{#1}}\,}
\def\gradgrad#1#2{\,\nabla\!_{{#1}}\nabla\!_{{#2}}\,}
\def\partder#1#2{{\partial #1\over\partial #2}}
\def\secder#1#2#3{{\partial^2 #1\over\partial #2 \partial #3}}
%
%
%
%
%
\def\ltap{\ \raise.3ex\hbox{$<$\kern-.75em\lower1ex\hbox{$\sim$}}\ }
\def\gtap{\ \raise.3ex\hbox{$>$\kern-.75em\lower1ex\hbox{$\sim$}}\ }
\def\gl{\ \raise.5ex\hbox{$>$}\kern-.8em\lower.5ex\hbox{$<$}\ }
\def\roughly#1{\raise.3ex\hbox{$#1$\kern-.75em\lower1ex\hbox{$\sim$}}}
%
%
%
%
\def\slash#1{\rlap{$#1$}/} 
\def\dsl{\,\raise.15ex\hbox{/}\mkern-13.5mu D} 
\def\delsl{\raise.15ex\hbox{/}\kern-.57em\partial}
\def\Ksl{\hbox{/\kern-.6000em\rm K}}
\def\Asl{\hbox{/\kern-.6500em \rm A}}
\def\Dsl{\hbox{/\kern-.6000em\rm D}} 
\def\Qsl{\hbox{/\kern-.6000em\rm Q}}
\def\gradsl{\hbox{/\kern-.6500em$\nabla$}}
%
%
\let\al=\alpha
\let\be=\beta
\let\ga=\gamma
\let\Ga=\Gamma
\let\de=\delta
\let\De=\Delta
\let\ep=\varepsilon
\let\ze=\zeta
\let\ka=\kappa
\let\la=\lambda
\let\La=\Lambda
\let\del=\nabla
\let\si=\sigma
\let\Si=\Sigma
\let\th=\theta
\let\Up=\Upsilon
\let\om=\omega
\let\Om=\Omega
\def\ph{\varphi}
%
%
%
\newdimen\pmboffset
\pmboffset 0.022em
\def\oldpmb#1{\setbox0=\hbox{#1}%
 \copy0\kern-\wd0
 \kern\pmboffset\raise 1.732\pmboffset\copy0\kern-\wd0
 \kern\pmboffset\box0}
\def\pmb#1{\mathchoice{\oldpmb{$\displaystyle#1$}}{\oldpmb{$\textstyle#1$}}
	{\oldpmb{$\scriptstyle#1$}}{\oldpmb{$\scriptscriptstyle#1$}}}
%
%
%
%
%
\def\bar#1{\overline{#1}}
\def\vev#1{\left\langle #1 \right\rangle}
\def\bra#1{\left\langle #1\right|}
\def\ket#1{\left| #1\right\rangle}
\def\abs#1{\left| #1\right|}
\def\vector#1{{\vec{#1}}}
\def\inv{^{\raise.15ex\hbox{${\scriptscriptstyle -}$}\kern-.05em 1}}
\def\pr#1{#1^\prime}  
\def\lbar{{\lower.35ex\hbox{$\mathchar'26$}\mkern-10mu\lambda}} 
\def\e#1{{\rm e}^{^{\textstyle#1}}}
\def\ee#1{\times 10^{#1} }
\def\om#1#2{\omega^{#1}{}_{#2}}
\def\imp{~\Rightarrow}
\def\coker{\mathop{\rm coker}}
\let\p=\partial
\let\<=\langle
\let\>=\rangle
\let\ad=\dagger
\let\txt=\textstyle
\let\h=\hbox
\let\+=\uparrow
\let\-=\downarrow
\def\dot{\!\cdot\!}
\def\vfilll{\vskip 0pt plus 1filll}
%

\begin{titlepage}
\begin{flushright}
{\large
IASSNS-HEP-96/20\\
hep-ph/9603212\\
February 1996\\
}
\end{flushright}
\vskip 2cm
\begin{center}
{\Large\bf Leptophobic $U(1)$'s and the $R_b$--$R_c$ Crisis\footnote{Research
supported in part by DOE grant DE-FG02-90ER40542, and by the Monell and
W.~M.~Keck Foundations. Email: {\tt babu@sns.ias.edu, kolda@sns.ias.edu,
jmr@sns.ias.edu.}}}
\vskip 1cm
{\large
K.S.~Babu,
Chris Kolda,
and John March-Russell}\\
\vskip 0.5cm
{\large\sl School of Natural Sciences,\\
Institute for Advanced Study,\\
Princeton, NJ~08540\\}
\end{center}
\vskip .5cm
\begin{abstract}
In this paper, we investigate the possibility of explaining both the $R_b$
excess and the $R_c$ deficit reported by the LEP experiments through 
$Z$-$Z'$ mixing
effects. We have constructed a set of models consistent with a restrictive set
of principles: unification of the Standard Model (SM) 
gauge couplings, vector-like additional
matter, and couplings which are both generation-independent and leptophobic.
These models are anomaly-free, perturbative up to the GUT scale, and contain 
realistic mass spectra. Out of this class of models, we find three explicit
realizations which fit the LEP data to a far better extent than the
unmodified SM or MSSM and satisfy all other phenomenological constraints which
we have investigated. One realization, the $\eta$-model coming from $E_6$, 
is particularly attractive, arising
naturally from geometrical compactifications of heterotic string theory. 
This conclusion depends crucially on the inclusion of a $U(1)$ kinetic mixing
term, whose value is correctly predicted by renormalization group running
in the $E_6$ model given one discrete choice of spectra.
\end{abstract}
\end{titlepage}
\setcounter{footnote}{0}
\setcounter{page}{1}
\setcounter{section}{0}
\setcounter{subsection}{0}
\setcounter{subsubsection}{0}

\section{Introduction \& Principles} \label{sec:intro}

During the past six years the four experiments at LEP 
have provided an abundance of data supporting the Standard Model (SM)
of particle physics and its $SU(3)_c\times SU(2)_L\times U(1)_Y$ gauge group
structure. Until recently there has been no significant deviation pointing
to new sources of physics beyond the SM. However, within the last
two years there has been growing evidence that a discrepancy exists between
the predicted and measured widths for the $b$ and $c$-quark decays of
the $Z$ boson. In particular, LEP has reported measurements of~\cite{lep}:
\beq
\left.\begin{array}{c} R_b \\ R_c \end{array}\right\rbrace
\equiv\frac{\Gamma(Z\to\bar bb/\bar cc)}{\Gamma(Z\to{\rm hadrons})}=
\left\lbrace\begin{array}{c}0.2219\pm0.0017 \\ 0.1543\pm0.0074\end{array}
\right.
\eeq
These values differ from the SM predictions, $R_b = 0.2152 \pm 0.0005$ and 
$R_c = 0.1714 \pm 0.0001$ \cite{Z0pole} 
(for $m_t=(176\pm 13)\gev$\cite{CDFtop} and $\alpha_s=0.125\pm0.010$), 
by $3.9\sigma$ and
$-2.3\sigma$ respectively.

If one is willing to accept the $R_c$ discrepancy as statistical, then there
are many new sources of physics which can serve to resolve the $R_b$ 
measurement by only changing the couplings of the third-generation fermions.
Such a method is naturally provided by low-energy supersymmetry (SUSY) 
with light charginos and stops~\cite{SUSYRb}, or by additional 
fermions mixing with, or additional interactions of, the $b$ and $t$ 
quarks~\cite{mixingRb}. However, if one interprets the $R_c$ deficit as
another signal of new physics, then the scenarios for new physics
are more limited~\cite{otherRbRc}.

A potential hurdle which one must face with respect to simultaneously
explaining the $R_b$ excess and the $R_c$ deficit is that
the LEP measurement for the total hadronic width of the $Z$ is in good
agreement with the SM prediction ($\Gamma_{\rm had}=(1744.8\pm3.0)\mev$ at
LEP versus $\Gamma_{\rm had}=(1743.5\pm3.1)\mev$ in the SM), while 
the sum $R_b+R_c$ is in slight disagreement with the SM prediction.
That is, $R_b+R_c=0.3762\pm0.0070$ as measured at LEP (with the error
correlations properly included), versus a theoretical expectation of
$0.3866\pm0.0005$, $1.5\sigma$ apart. 

A clue to solving this conundrum may lie in a simple observation. Defining
$\De\Gamma_i$ as the difference between the experimental and the theoretical
determinations of $\Gamma_i$, one notes that 
\beq
3\Delta\Gamma_b+2\Delta\Gamma_c=(-23.2\pm24.3)\mev
\eeq
so that at the $1\sigma$ level, a consistent interpretation of the data is 
given by assuming a flavor-dependent but generation-{\sl independent} shift
in the hadronic $Z$-couplings. That is, 
\bea
\Gamma_{u,c}&=&\Gamma^{\rm SM}_{u,c}+\Delta\Gamma_c \nonumber \\
\Gamma_{d,s,b}&=&\Gamma^{\rm SM}_{d,s,b}+\Delta\Gamma_b. \label{eq:genindep}
\eea
Such a pattern of shifts has also been suggested in~\cite{feng,altar,chia}.

A second hurdle in explaining the $R_b$ and $R_c$ puzzles is that
unlike the partial hadronic widths of the $Z$, the well-measured
partial leptonic widths are 
in good agreement with the SM predictions: $\Gamma_e=83.93\pm0.14\mev$
and $\Gamma_{\rm inv}=499.9\pm2.5\mev$, which are within $0.4\sigma$ and 
$-0.4\sigma$ respectively of theory. Any source of new physics must 
preserve the successful predictions of the SM for the leptonic widths.

In this paper we propose to explain the $R_b-R_c$ problem by introducing
an additional $U(1)'$ gauge symmetry. If this new $U(1)'$ is broken
near the electroweak scale, there can be significant mixing between the usual
$Z$ and the new $Z'$. The physical $Z$-boson as produced at LEP will then have
its couplings to fermions altered by an amount proportional to the
$Z-Z'$ mixing angle times the $Z'$ coupling to those same fermions. 

Analyses have recently appeared in the literature~\cite{altar,chia} 
that seek to fit the LEP data by introducing such an additional $U(1)'$.
Both of these works make a phenomenological fit to the data
introducing some number of new parameters, such as arbitrary $U(1)'$
charge ratios, $Z-Z'$ mixing angle, and $Z'$ mass.
These analyses do indicate that this class of scenarios
has the potential to solve the $R_b-R_c$ discrepancy, and are therefore
interesting. However, they share some
fundamental problems associated with the lack of an underlying,
consistent framework. For example, the extra $U(1)'$ is {\sl not}
anomaly free (this is true both for the $[U(1)']^3$, and most
seriously, the mixed SM-$U(1)'$ anomalies). Further, 
since the authors of~\cite{altar,chia} also seek to explain
the CDF dijet excess, they are forced to take a high value of
the $Z'$ mass. For such $Z'$ masses, the $U(1)'$-couplings
have to be so large that the $U(1)'$ gauge coupling becomes
non-perturbative at most a decade above the $Z'$ mass scale;
implicit in this is that the $Z'$ width in these models equals or
even exceeds the $Z'$ mass. 

Here we will take a different approach.  We set
forth a few basic principles which we believe any
attractive $Z'$-model should obey.  Within this framework 
we will find that there exist only limited classes of $U(1)'$
models which are phenomenologically viable and theoretically consistent.
Each class has a well-defined 
prediction for the $U(1)'$ charges of the SM fermions, reducing much of
the arbitrariness in the couplings. We will not attempt to explain the
CDF dijet anomaly.

The principles that we demand are:
\begin{itemize}

\item The low energy spectrum must be consistent with the unification of
the {\it standard model} gauge couplings that occurs in the minimal
supersymmetric standard model (MSSM).
This will lead us to consider models which are extensions of the
MSSM, with any non-MSSM matter added in particular combinations which
can be thought of as filling complete multiplets of $SU(5)$. We
allow the possibility of unification within a string framework, and
do not {\it require} the presence of a field theoretic GUT.

\item All non-MSSM matter must fall into vector-like representations under
the SM gauge groups. Such a requirement is consistent with the absence of
experimental evidence for new fermions with masses below the top quark mass.
Further, note that additional {\sl chiral} matter is disfavored by the
electroweak precision measurements, since,
in contrast to vector-like matter it can give very large contributions to
the $S$, $T$, and $U$ parameters.

\item The $U(1)'$ charges of the SM leptons must be (to a good approximation)
zero. This requirement of {\sl leptophobia} is motivated by the
phenomenology. This alone will eliminate the $U(1)$ factors
associated with most traditional GUT groups, since GUT's tend to place 
leptons and quarks into common multiplets. 

\item Consistent with Eq.~(\ref{eq:genindep}), we require that the
$U(1)'$ couplings be generation-independent. This requirement is essential
if tree--level hadronic flavor changing neutral current processes mediated
by the $U(1)'$ gauge boson are to be naturally suppressed.  
This also has the advantage of simplicity and economy. 

\end{itemize}

To be precise, the principle of unification that we will impose requires
that the meeting of the SM couplings at $2\times10^{16}~\gev$ is not
a coincidence. For simplicity we will not explicitly
consider in this article 
the various string models where the scale of unification is increased
to the (weak-coupling prediction of the) string unification scale
$M_{\rm str}^{\rm 1-loop}\sim 5\times 10^{17}~\gev$, such as those
discussed in~\cite{stringuni}, although it will be clear
that the consequences for our discussion of such a modification are
slight. (Note that one interesting possibility that could maintain unification 
at $2\times10^{16}~\gev$ is the strongly coupled string scenario
recently proposed by Witten~\cite{witten}.) 

If one takes the unification of gauge couplings to imply the existence
of a simple GUT gauge group, then the natural candidates with extra $U(1)$'s
and three chiral families are $SO(10)$ and $E_6$. However the single additional
$U(1)$ within $SO(10)$ is not leptophobic.
In $E_6$ all linear combinations 
of the two additional $U(1)$'s orthogonal to hypercharge 
couple to leptons. Nonetheless, we will show that by including an effect
usually overlooked in the literature ($U(1)$-mixing in the kinetic terms
through renormalization group flow~\cite{holdom,delarge}) there exists
a unique $U(1)'$ in the $E_6$ group which is compatible with the
data. The $E_6$ subgroup in question is usually known in the literature as
the $\eta$-model and interestingly is the unique model which results from
$E_6$ Wilson--line breaking directly to a rank-5 subgroup in a string
context~\cite{E6review}. We will discuss this case in
some detail in Section~\ref{sec:etamodel}.

Finally, although we assume the MSSM for the purposes of gauge coupling 
unification, we do not use MSSM loop contributions to the $Z\bar bb$
vertex in order to explain any part of the $R_b$ anomaly. In particular
we do not assume light charginos or top squarks which are the necessary
ingredients for such a scenario~\cite{SUSYRb}.

\section{$Z-Z'$ Mixing} \label{sec:zmix}

We begin with a brief general discussion of $Z-Z'$ mixing in the context of an
$SU(2)_L\times U(1)_Y\times U(1)'$ model. A more detailed discussion can
be found, for example, in Refs.~\cite{zmix,holdomrho}.
The neutral current Lagrangian of the $Z$ and $Z'$ is given by
\beq
\CL_{\rm NC}=\frac{1}{2}\sum_i\bar\psi_i\gamma^\mu\left(\frac{g_2}{\cw}
(v_i+a_i\gamma^5)Z_\mu + g'(v'_i+a'_i\gamma^5)Z'_\mu\right)\psi_i
\label{eq:newLnc}
\eeq
where 
\beq
v_i=T_{3i}-2Q_i\swsq, \quad\quad\quad\quad a_i=-T_{3i}
\eeq
are the SM vector and axial couplings of the $Z$, and $v'$, $a'$ are the
(unknown) vector and axial couplings of the $Z'$. Here $g'$ is the coupling
constant of the new $U(1)'$ and $\swsq\equiv\sin^2\theta_W$.

After electroweak and $U(1)'$ breaking, the $Z$ and $Z'$ gauge bosons mix
to form the mass eigenstates $Z_{1,2}$, where we will identify the
$Z_1$ with the gauge boson produced at LEP:
\bea
Z_1&=&\cos\xi\, Z + \sin\xi\, Z' \nonumber \\
Z_2&=&- \sin\xi\, Z + \cos\xi\, Z' .
\eea
Since such mixing must necessarily be small in order to explain the general
agreement between LEP results and the SM, we will throughout this paper use the
approximation $Z_1\simeq Z + \xi Z'$. We will also assume that
the mass of the $Z_2$ is large enough so that its effects at LEP, either
via direct production or loop effects can be ignored. Therefore
all new physics effects must appear through the mixing angle $\xi$. The 
relevant Lagrangian probed at LEP will then be
\beq
\CL_{Z_1}=\frac{g_2}{2\cw}\sum_i\bar\psi_i\gamma^\mu(\bar v_i+\bar a_i\gamma^5)
Z_{1\mu}\psi_i
\eeq
where, for small $\xi$,
\bea
\bar v_i&\simeq&v_i + \bar\xi v'_i \nonumber \\
\bar a_i&\simeq&a_i + \bar\xi a'_i, \label{eq:couples}
\eea
and we have defined the auxiliary quantity
\beq
\bar\xi\equiv(g'\cw/g_2)\,\xi. 
\label{eq:xibar}
\eeq

Because the $Z_1$ is no longer purely the electroweak $Z$, the
$\rho$-parameter
\beq
\rho-1\equiv4\sqrt{2}\,G_F\left(\Pi_{11}(0)-\Pi_{33}(0)\right) 
\label{eq:defrho}
\eeq
receives a tree-level correction. (Here $\Pi_{ii}(0)$ are the $SU(2)_L$ 
vacuum polarization amplitudes at zero momentum transfer.)
If we define the corrections to $\rho$
by
\beq
\rho\equiv1+\De\rho_{SM}+\De{\bar\rho},
\eeq
where $\De\rho_{SM}$ is due to loop corrections already
present in the SM (such as the top), then the mixing with the
$Z'$ contributes to $\De{\bar\rho}$. Since we will later be interested
in taking into account the effects of further shifts in $\rho$ due to
the rest of the MSSM spectrum, we decompose
$\De{\bar\rho}=\De\rho_{\rm M} + \De\rho_{\rm extra}$, where $\De\rho_{\rm M}$
is the part due to mixing with the $Z'$. The value of $\De{\bar\rho}$
is the quantity that our fits to the LEP data will directly constrain. 
Writing the $Z-Z'$ mass matrix as
\beq
M^2_{Z,Z'}=\left(\begin{array}{cc} m^2_Z & \Delta m^2 \\ \Delta m^2 & M^2_{Z'}
\end{array}\right), \label{eq:mzz}
\eeq
then for $M^2_{Z'}\gg m^2_Z$, $\Delta m^2$, one finds that the shift in $\rho$
due to mixing, $\Delta\rho_M$, is given by
\beq
\Delta\rho_M\simeq\xi^2\left(\frac{m^2_{Z_2}}{m^2_{Z_1}}\right)
\simeq\xi^2\left(\frac{M^2_{Z'}}{m^2_{Z}}\right), 
\label{eq:delrhoM}
\eeq
where
\beq
\xi \simeq -{\Delta m^2 \over M_{Z'}^2}.
\label{eq:xi}
\eeq
There is also a corresponding shift in $\swsqb$:
\beq
\swsqb=\swsqb|_{\xi=0} - \frac{s_W^2c_W^2}{c_W^2-s_W^2}{\Delta\rho_M}.
\label{eq:swsqb} 
\eeq
In terms of the above parameters, one can then calculate the $Z_1$ partial
width to fermions:
\beq
\Gamma(Z_1\to\bar ff)=\frac{G_F\,m^3_{Z_1}}{6\sqrt{2}\,\pi}\rho N_c\left(
\bar v_f^2+\bar a_f^2\right).
\eeq

A further relation may be obtained by examining the specific form of the terms
that come into Eq.~(\ref{eq:mzz}). If we assume that the fields $\phi_i$ which
receive vev's occur only in doublets or singlets of $SU(2)_L$, then
\bea
m^2_Z&=&\frac{2g^2_2}{\cwsq}\sum_i\langle T_{3i}\phi_i\rangle^2\quad
=\quad\frac{g^2_2}{2\cwsq}v^2_Z, \nonumber \\
M^2_{Z'}&=&2g'^2\sum_i\langle Q'_i\phi_i\rangle^2, \\
\Delta m^2&=&\frac{2g_2 g'}{\cw}\sum_i\langle T_{3i}\phi_i\rangle
\langle Q'_i\phi_i\rangle, \nonumber
\eea
where $Q'_i$ is the $U(1)'$ charge of $\phi_i$ and $v^2_Z$ is the sum of the
vev's of the $SU(2)_L$ doublets. Then we may write
$\Delta\rho_M$ as a simple function of $\bar\xi$:
\beq
\Delta\rho_M\simeq-\left(\frac{g_2}{g'\cw}\right)\left(\frac{\Delta m^2}
{m^2_Z}\right)\bar\xi = 
-{4\bar{\xi}\over v^2_Z}\sum_i\langle T_{3i}\phi_i\rangle
\langle Q'_i\phi_i\rangle.
\label{eq:rhoxi}
\eeq
What is noteworthy about this relationship is that it is connects the two
quantities ($\Delta\rho_M$ and $\bar\xi$) which are experimentally
constrained at LEP (up to $\De\rho_{\rm extra}$, which we can bound), 
in a way that is independent of the unknown gauge
coupling $g'$ and the $Z'$ mass. Note that $\Delta m^2$ and $\bar{\xi}$
in Eq.~(\ref{eq:rhoxi}) have opposite signs, so that $\Delta \rho_M$ is
always positive.  

\subsection{$U(1)_a\times U(1)_b$ Mixing and RGE's} \label{sec:rge}

The discussion so far has echoed the conventional wisdom on the subject of
$Z-Z'$ mixing. However, it was realized many years ago~\cite{holdom}\ that in
a theory with two $U(1)$ factors, there can appear in the Lagrangian a term
consistent with all gauge symmetries which mixes the two $U(1)$'s. In the
basis in which  the interaction terms have the canonical form, the
pure gauge part of the Lagrangian for an arbitrary $U(1)_a\times U(1)_b$
theory can be written
\bea
\CL&=&-\frac{1}{4}\,F_{(a)}^{\mu\nu}F_{(a)\mu\nu}
-\frac{1}{4}\,F_{(b)}^{\mu\nu}F_{(b)\mu\nu}
-\frac{\sin\chi}{2}\,F_{(a)}^{\mu\nu}F_{(b)\mu\nu} \nonumber \\
& & +~\De m^2 A_{(a)\mu} A_{(b)}^\mu 
+\frac{1}{2}m_a^2 A_{(a)\mu} A_{(a)}^\mu
+\frac{1}{2}m_b^2 A_{(b)\mu} A_{(b)}^\mu
\eea
If both
$U(1)$'s arise from the breaking of some simple group $G\to U(1)_a\times
U(1)_b$, then sin$\chi=0$ at tree level. However, if the matter of the
effective low-energy supersymmetric theory is such that
\beq
\sum_{i={\rm chiral~fields}} \left(Q_a^i Q_b^i \right) \neq 0,
\eeq
then non-zero $\chi$ will be generated at one-loop.
This is naturally the case when split multiplets of the original non-Abelian
gauge symmetry, such as the Higgs doublets in a grand unified theory,
are present in the effective theory. Since we are interested in a
large separation of scales, $M_{GUT}$ and $M_Z$, we will need to resum
the large logarithms that appear \cite{delarge,delaguila} using the 
renormalization group equations (RGE's) for the evolution of the gauge
couplings including the off-diagonal terms.  

Once a non-zero $\chi$ (or $\De m^2$) has been induced, one needs to 
transform to the mass eigenstate basis. To do so, one must
perform a (non-unitary) transformation on the original gauge fields,
$A_{(a)}$ and $A_{(b)}$, to arrive at the mass eigenstates,
$Z_{1,2}$:
\bea
A_{(a)}&=&\left(\cos\xi -\tan\chi\sin\xi\right)Z_{1}
-\left(\sin\xi+\tan\chi\cos\xi\right)Z_{2} \nonumber \\
A_{(b)}&=&\left(\sin\xi\,Z_{1}+\cos\xi\,Z_{2}\right)/\cos\chi,
\label{eq:transf}
\eea
where
\beq
\tan2\xi=\frac{-2\cos\chi\,(\De m^2-m_a^2\sin\chi)}
{m_b^2-m_a^2\cos2\chi+2\,\De m^2\sin\chi}. \label{eq:mixangle}
\eeq
This transformation results in a shift in the effective charge to which one
of the original $U(1)$'s couples. (One $U(1)$ can always be chosen to have 
unshifted charges.) This can be seen by taking the $\xi=0$ limit of the 
above transformation. The resulting interaction Lagragian is 
then of the form~\cite{holdom}:  
\beq
\CL_{\rm int}=\bar\psi\,\gamma^\mu\left(g_aQ_aZ_{1\mu}+(g_bQ_b+g_{ab}Q_a)
Z_{2\mu}\right)\psi
\label{eq:Lmix}
\eeq
where the redefined gauge couplings are related to the original
couplings, $g^0$, by $g_a = g_a^0$, $g_b = g_b^0/\cos\chi$ and $g_{ab}=
-g_a^0\tan\chi$.
The ratio $\de\equiv g_{ab}/g_b$ is
a phenomenologically useful parameter, representing the shift in the
$Z_2$--fermion coupling due to kinetic mixing.

The renormalization group equations for the coupling--constant flow
of a $U(1)_a\times U(1)_b$ theory, including the off--diagonal mixing,
are most usefully formulated in the basis of Eq.~(\ref{eq:Lmix}).
In this basis the
equations for the couplings $g_a, g_b$ and $g_{ab}$ are:
\bea
{dg_a\over dt}&=&\frac{1}{16\pi^2}g_a^3 B_{aa}, \nonumber \\
{dg_b\over dt}&=&\frac{1}{16\pi^2}g_b\biggl( g_b^2 B_{bb} + g_{ab}^2 B_{aa}
 + 2g_b g_{ab}B_{ab} \biggr), \label{eq:rge} \\
{dg_{ab}\over dt}&=&\frac{1}{16\pi^2}\biggl(g_b^2 g_{ab}B_{bb} + g_{ab}^3
B_{aa} + 2g_a^2 g_{ab}B_{aa} + 2g_a^2g_b B_{ab} + 2g_b g_{ab}^2 B_{ab}\biggr),
\nonumber 
\eea
where $B_{ij} = \tr(Q_i Q_j)$ with the trace taken over all the chiral
superfields in the effective theory, and there is no sum over
$(a,b)$ in Eq.~(\ref{eq:rge}). From these
equations we immediately see that even if $g_{ab}=0$ to begin with, a
non-zero value of the off-diagonal coupling is generated if the inner--product
$\tr(Q_i Q_j)$ between the two charges is non--zero. The advantage of this
basis for the RGE's is that the low-energy value of the parameter $\de$
is given directly by the ratio $g_{ab}/g_b$ evaluated at the low scale.
(This is not the case for the more symmetrical form of the RGE's given
in Ref.~\cite{delarge}.) 

For the case at hand, we will choose the couplings of 
the usual $Z_\mu$ to be canonical, shifting the charge of the $Z'_\mu$. 
Since it is the $B_\mu$ component of $Z_\mu$ which mixes through the kinetic
terms, the couplings
of the $Z'$ to matter fields can be expressed in terms of an effective
$U(1)'$ charge $\Qeff=Q'+Y\delta$, where $Y$ is hypercharge.
We can translate from Eq.~(\ref{eq:Lmix}) using $g_a=g_2/\cw$ and
$g_b=g'$ so that $g_{ab}=-g_2\tan\theta_W\tan\chi$ and $\delta=g_{ab}/g_b$.
The vector and axial couplings that come into Eq.~(\ref{eq:couples})
are given by
\bea
v'&=&\Qeff(\psi) - \Qeff(\psi^c) \nonumber \\
a'&=&-\Qeff(\psi) - \Qeff(\psi^c).
\eea
Note that both $\psi$ and $\psi^c$ are left-handed chiral fields:
$\Qeff(\psi^c)=-\Qeff(\psi_R)$.

In most of the models we will consider, we will work directly with
$\Qeff$; in such models, whether or not $\Qeff$ can be expressed as
some $Q'+Y\delta$ for non--zero $\delta$ will not have an effect on the
analysis. However, when considering the $\eta$--model coming from $E_6$,
the difference between $\Qeff$ and $Q_\eta$ will have important
consequences on the observable physics. We reserve further comment on the
$U(1)$ mixing in the $E_6$ model until Section~\ref{sec:etamodel}.

Kinetic mixing of $U(1)$'s will also shift the 
$\rho$-parameter. In the previous subsection we had assumed that we
could write the electroweak $Z$ in terms of the mass eigenstates as
$Z=\cos\xi\,Z_1-\sin\xi\,Z_2$. 
However, in the presence of a non--zero $\chi$ (or
$\delta$), this is changed to (see Eq.~(\ref{eq:transf}), replacing
$\tan\chi$ with $-\sw\tan\chi$):
\bea
Z&=&(\cos\xi+\sin\xi\sw\tan\chi)\,Z_1-(\sin\xi-\cos\xi\sw\tan\chi)\,Z_2
\nonumber \\
Z'&=&(\sin\xi\,Z_1+\cos\xi\,Z_2)/\cos\chi \label{eq:wavefunctions}\\
A&=&\gamma-\cw\tan\chi(\sin\xi\,Z_1+\cos\xi\,Z_2) \nonumber 
\eea
where $\gamma$ is the physical photon. 
Eq.~(\ref{eq:mixangle}) for $\xi$ becomes
\beq
\tan2\xi=\frac{-2\cos\chi(\De m^2+m_Z^2\sw\sin\chi)}
{M^2_{Z'}-m_Z^2\cos^2\chi+m_Z^2\swsq\sin^2\chi-2\,\De m^2\sw\sin\chi},
\eeq
while the $Z_1$ mass is given to lowest order in $m_Z^2/M_{Z'}^2$ by:
\beq
m_{Z_1}^2=m_Z^2\left\lbrace 1-\frac{m_Z^2}{M_{Z'}^2}\left(\frac{\De m^2}
{m_Z^2}+\sw\sin\chi\right)^2\right\rbrace.
\eeq

The coefficient of the $Z_1$ term in Eq.~(\ref{eq:wavefunctions}) 
is essentially
a wave-function renormalization for the $Z_1$ and contributes to 
$\Delta\rho_M$ by absorbing part of the explicit mass shift which came from
mass matrix mixing~\cite{holdomrho}. The net effect is a {\sl negative}
contribution to $\Delta\rho_M$ which subtracts from the positive definite
contribution coming from mass mixing. In terms of $\delta$,
\beq
\Delta\rho_M\simeq\frac{M^2_{Z'}}{m_Z^2}\xi^2-2k\xi\delta
\label{eq:drho2}
\eeq
where $k=g'c_W s_W/g_2$. The important point
to note is that, in the presence of kinetic mixing, $\Delta\rho_M$ can be
smaller than had there been no such mixing; in fact, $\Delta\rho_M$ can be
negative.

The kinetic mixing also shifts $\swsqb$ beyond what was already included
in Eq.~(\ref{eq:swsqb}):
\beq
\swsqb=\swsqb|_{\xi=\delta=0} - \xi c_W^2\left(\frac{s_W^2}{c_W^2-s_W^2}
\frac{M^2_{Z'}}{m_Z^2}\xi + k\delta\right).
\label{eq:swsqb2} 
\eeq
For $\delta=0$ this reduces to Eq.~(\ref{eq:swsqb}). Finally, there is a
new contribution, $S_M$, to the so-called $S$-parameter (see, \eg,
Ref.~\cite{holdomrho}) due to kinetic mixing which can be negative:
\beq
\alpha S_M\simeq-4\cwsq k\xi\delta
\eeq
to leading order in $m_Z^2/M_{Z'}^2$.

\subsection{New Contributions to Oblique Parameters} \label{sec:oblique}

As noted in the previous Sections, in the absence of $U(1)$ kinetic
mixing (\ie, $\delta=0$),
$Z$--$Z'$ mixing gives a positive contribution
to the $\rho$--parameter, denoted by $\Delta \rho_M$, and no contribution
to the $S$-parameter.  
Since our numerical fits are sensitive to the total $\Delta \bar\rho$ and $S$, 
it is important to see if there are corrections from sources other than the 
$Z$--$Z'$ mixing. (Both $\De\bar\rho$ and $S$ are
defined to be zero in the SM for some reference top quark and Higgs boson 
masses which we take to be $175\gev$ and $125\gev$ respectively.)
The spectrum of the effective theory in all 
models that we will consider includes a Higgs sector with two doublets, 
vector--like states in complete ``$SU(5)$ multiplets'', and the superpartners
of all particles, all of which can in principle contribute to the oblique
parameters.
The sizes of these contributions depends on the details of the mass
spectrum. As we shall see, the scale of the $U(1)'$ breaking turns
out to be relatively low in all models (typically $M_{Z'}\sim 200 - 
250\gev$). Therefore the contributions of the additional matter cannot
be ignored in general.
Let us therefore estimate the typical allowed ranges for $\De\rho_{\rm extra}$
and $S_{\rm extra}$ ($S\equiv S_M+S_{\rm extra}$), 
given some reasonable choices for the spectrum, 
in particular that depending upon MSSM superpartners, Higgs sector, and
additional vector-like matter.

The superpartner
contributions to $\De\rho_{\rm extra}$ and $S_{\rm extra}$ in the MSSM 
have been studied in Refs.~\cite{drees}
and~\cite{haber}\ respectively. In Ref.~\cite{haber}\ it has been shown
that such contributions to $S_{\rm extra}$ are generally very small; 
therefore we will ignore MSSM superpartner 
contributions to $S_{\rm extra}$ in everything that follows.
Likewise it is shown in Ref.~\cite{drees}\ that the corrections to
$\De\rho_{\rm extra}$ from the MSSM sparticle spectrum are small 
(and positive) with
the exception of the stop--sbottom correction which can be sizable
depending on the nature of the supersymmetric spectrum. 

Although the Higgs--boson contribution to $\De\rho_{\rm extra}$
in a general two--doublet model can be
large and negative (as large as $-0.01$), in supersymmetric
models there are restrictions on the Higgs sector parameters, resulting in 
an absolute lower bound of $\Delta\rho_{\rm extra}\geq-0.0015$ from the MSSM
Higgs sector. However, in the class of models 
which we will consider
in Section~\ref{sec:models}, this number becomes $-0.002$ since
the Higgs sector in these models is not identical to that of the MSSM. This
is because the $\mu H_u H_d$ term of the MSSM will be replaced by $\lambda
H_u H_d S$, where $S$ is a SM--singlet field carrying $U(1)'$ charge.  
There is also a new contribution to the Higgs potential from
the $U(1)'$ D--term.  We have analyzed the Higgs spectrum of these models,
which resemble the MSSM with a singlet (the NMSSM).
In the limit where the singlet vev
is large compared to the doublet vev's, but keeping the mass of the
pseudoscalar fixed, we have numerically examined the most negative
$\Delta \rho_{\rm extra}$
obtainable from the Higgs sector and found it to be $-0.002$.  Of course, this 
could be partially offset by some positive contribution from other sectors, 
such as the stop-sbottom sector. In the model analysis of 
Section~\ref{sec:expts} we will therefore consider two cases, one in which
$\De\rho_{\rm extra}=0$ and another in which we take $\De\rho_{\rm extra}$
to have the not unreasonable value $-0.001$.

As far as the contributions from additional vector-like matter are concerned,
we will always consider the simple isospin-symmetric case 
(\ie, the masses of the $T_3=\pm \frac{1}{2}$ states equal) where 
there are no vector-like contributions to $\De\rho_{\rm extra}$. In this
limit, $S_{\rm extra}$ need not be zero. 
For the various models we will consider, $S_{\rm extra}$ 
receives potentially large contributions from the multiplicity of 
lepton/higgsino 
doublets which arise. There are two natural cases. One, in which
the vector-like contributions to the doublet masses dominate over the
chiral contributions, gives $S_{\rm extra}\simeq0$. Alternatively,
because the weak scale and the $U(1)'$ scale are quite close, the chiral
masses can be of order the vector-like masses;
we have estimated, using the results of Ref.~\cite{maroy}, the contribution 
to $S_{\rm extra}$ in this case to be $+0.14$ per pair of such doublets.

\section{Leptophobic U(1) Models} \label{sec:models}

Any model which hopes to extend the SM in a minimal fashion must give masses
to the SM fermions through the usual Higgs mechanism. Within a supersymmetric
model, such couplings appear in the superpotential, $W$. Letting $W_0$ be
the minimal superpotential consistent with the SM, we write\footnote{
With the extended matter content that we will introduce
later in the paper, it is also possible to consider more complicated
non-minimal choices for these Yukawa couplings, where the Higgs
that couples to $e^c$ and $d^c$ are distinct. We will not
analyze these possibilities in detail here.}
\beq
W_0=h_uQH_uu^c + h_dQH_dd^c + h_eLH_de^c.
\eeq
The new $U(1)'$ must also preserve this superpotential. Demanding that the 
$U(1)'$ couplings of the leptons be zero allows us to write the charges
of the remaining fields as:
\beq \begin{array}{ll}
Q'(Q)\equiv x & Q'(H_u)=-x-y \\
Q'(u^c)\equiv y & Q'(H_d)=0 \\
Q'(d^c)=-x\quad\quad & {} \end{array} \label{eq:charges}
\eeq

We next require that the resulting gauge theory have no anomalies. In the case
of the SM particle content alone, this implies $C_3=C_2=C_1=C_0=0$, where,
\bea
\lbrack SU(3)\rbrack{}^2\times U(1)': & 3x+3y &\equiv C_3 \label{eq:31} \\
\lbrack SU(2)\rbrack{}^2\times U(1)': & 8x-y  &\equiv C_2 \label{eq:21} \\
\lbrack U(1)_Y\rbrack{}^2\times U(1)': & -x+\frac{7}{2}y &\equiv C_1 
\label{eq:Y1} \\
\lbrack U(1)'\rbrack{}^2\times U(1)_Y: & (x+y)(7x-5y) &\equiv C_0.
\label{eq:1Y}
\eea
At this time we do not concern ourselves with the $[U(1)']^3$, or
$U(1)'[{\rm gravity}]^2$ anomalies since these can be saturated with
any number of SM gauge singlets. The only solution which 
cancels all anomalies in Eqs.~(\ref{eq:31})--(\ref{eq:1Y}) is the trivial
solution $x=y=0$.

Going beyond the MSSM, we wish to add matter in such a way that the 
unification of gauge couplings that occurs in the MSSM is not upset. To do
so we must arrange that the additional matter changes the MSSM one--loop
beta--function coefficients in such a way that
$\Delta b_2=\Delta b_3={3\over 5}\Delta b_1$. This constraint can be
most easily understood as requiring the addition of complete $SU(5)$ multiplets
to the spectrum (though $U(1)'$ need not commute with this fictitious
$SU(5)$). 

Our principles outlined in Section~\ref{sec:intro}\ constrain us further in 
how we add
$SU(5)$ multiplets to the model. Implicit in the requirement of unification
is that the gauge couplings remain perturbative up to the unification scale.
This implies that we can only add (a limited number of) ${\bf5}$'s,
${\bf10}$'s, and their conjugate
representations. By requiring that all new matter be
vector-like under the SM gauge groups, we restrict ourselves further to
adding the multiplets in pairs. In combination, these two principles limit
us to adding (A) up to four $(\fpf)$ pairs, or (B) one $(\tpt)$ pair,
or (C) one pair each of $(\fpf)$ and $(\tpt)$.

Consider Model A with a single pair of $(\fpf)$. Because we require neither
that the $U(1)'$ commutes with the ersatz $SU(5)$, nor that the charge
assignments be vectorial with respect to the $U(1)'$, we write general
$U(1)'$ charges for the new states as:
\beq \begin{array}{ccccc}
{\bf5}&=&({\bf3},{\bf1})\,[-1/3,a_1]&+&({\bf1},{\bf2})\,[1/2,a_2] \\
{\bf\bar5}&=&({\bf\bar3},{\bf1})\,[1/3,\bar a_1]&+&({\bf1},{\bf2})\,
[-1/2,\bar a_2]
\end{array}\label{eq:fpfcharges}
\eeq
where each state is listed by its $(SU(3)_c,SU(2)_L)\,[U(1)_Y,U(1)']$
representation/charge. The anomaly coefficients are changed to:
\beq
\begin{array}{ll}
C_0\to C_0 -a_1^2+a_2^2+\bar a_1^2-\bar a_2^2 & C_2\to C_2 + a_2+\bar a_2 \\
C_1\to C_1 +\frac{1}{3}(a_1+\bar a_1)+\frac{1}{2}(a_2+\bar a_2)\quad &
C_3\to C_3 +a_1+\bar a_1.
\end{array}
\eeq
Solving for the condition $C_3=C_2=C_1=C_0=0$ yields 
\beq
y=2x,
\label{eq:Acharge}
\eeq
with the additional relations $a_1 = -2(\bar a_2 + 9x)/3$,
$a_2=-\bar a_2-6x$, and
$\bar a_1=(2\bar a_2-9x)/3$. Note that all charges are rationally related, 
and, further, that for a purely axial choice of $U(1)'$ charges
($a_1=\bar a_1$ \etc), the only solution is the trivial one $x=y=a_i=0$.
The result Eq.~(\ref{eq:Acharge})
does not depend on the number of $(\fpf)$ pairs. Thus for this entire class
of models, we know the couplings of all the quarks to the $Z'$ through
Eq.~(\ref{eq:charges}), up to one overall normalization. 

The same exercise can be undertaken for Model B. Now we add the states in the
$(\tpt)$ with charge assignments
\beq \begin{array}{ccccccc}
{\bf10}&=&({\bf3},{\bf2})\,[1/6,a_3]&+&({\bf\bar3},{\bf1})\,[-2/3,a_4]
&+& ({\bf1},{\bf1})\,[1,a_5] \\
{\bf\bar10}&=&({\bf\bar3},{\bf2})\,[-1/6,\bar a_3]&+&
({\bf3},{\bf1})\,[2/3,\bar a_4] &+& ({\bf1},{\bf1})\,[-1,\bar a_5]. 
\end{array}\label{eq:tptcharges}
\eeq
In the general case the phenomenologically important ratio $y/x$ is
undetermined by the
anomaly conditions. However, if we make the very natural simplifying
assumption that the $U(1)'$ charges in Eq.~(\ref{eq:tptcharges}) are
purely axial ($a_3 =\bar a_3$, \etc), then the
$[U(1)']^2\times U(1)_Y$ anomaly equation (\ref{eq:1Y}) is unmodified
and there are only two solutions for the charge ratio: 
\beq
y=-x,\quad {\rm or}\quad y={7x\over 5}.
\label{eq:Bcharge}
\eeq
The associated charges of the extra states are $\{a_3,a_4,a_5\}=
\{-3x/2,3x,-3x/2\}$ and $\{-11x/10,-7x/5,x/10\}$ respectively.
In the following we will refer to these models as ``B(-1)'' and ``B(7/5)''.
In the ``B(-1)'' model the charges are identical to baryon number, with the
Higgs doublet $H_u$ carrying zero charge.   
At this stage it is
important to recognize that both these models have the potential
problem that the extra states do not include
$({\bf 1},{\bf 2})_{\pm1/2}$ representations which
can be used to give a naturally small off-diagonal mixing term $\De m^2$ in
the $M^2_{Z,Z'}$ mass matrix Eq.~(\ref{eq:mzz}).  In the B(-1) model, there is
no tree--level $Z-Z'$ mixing.  Even at the one--loop level, no such mixing
arises in the simplest version of this model where the $({\bf 10} + 
{\bf \bar{10}})$ states receive masses from SM singlets only.  In the
B(7/5) model, on the other hand, there is tree--level $Z-Z'$ mixing,
which however tends to be too large.   As we will see, this
model requires additional (negative) contributions to the
$\rho$-parameter to relax the constraint Eq.~(\ref{eq:rhoxi}).  

Model C has, in the general case, ten new $U(1)'$ charges corresponding
to the ten new states in Eqs.~(\ref{eq:fpfcharges}) and (\ref{eq:tptcharges}),
and again even with the constraints imposed by anomaly cancellation the ratio
$y/x$ is not determined. However there are two particularly attractive
and natural subclasses of these models. In the first subclass the $U(1)'$
charges of the extra states are chosen to be purely axial. This leads
to the charge ratios $y/x=-1$ or $7/5$ as in Eq.~(\ref{eq:Bcharge}) (Models
``C(-1)'' and ``C(7/5)'' respectively). Note that since all C-type models
contain an extra pair of Higgs doublets, they are 
naturally able to accommodate a suitably small $Z-Z'$ mixing. The second
attractive subclass of Model~C is defined by setting the $U(1)'$ charges
of the anti-generation (${\bf 5} + {\bf \bar 10}$) to zero
($a_1=a_2=\bar a_3=\bar a_4=\bar a_5 =0$). In this case 
the ratio $y/x$ is continuously adjustable as is the charge, $a_3$, of the
additional $({\bf 3},{\bf 2})_{1/6}$ state. Among this continuous
family, the choice
\beq
y=x
\label{eq:Ccharge}
\eeq
is especially simple and attractive (Model ``C(1)''). 

In all cases we still need to impose the $[U(1)']^3$ and
$U(1)'[{\rm gravity}]^2$ anomaly cancellation conditions. It
is important to consider an efficient way of achieving this because 
we will soon see that there is a strong constraint arising from the
requirement of perturbativity of the $U(1)'$-coupling all the way up to
the GUT scale, and the $U(1)'$ beta--function gets a significant 
contribution from these SM-singlet states (collectively $\Si$'s).
One must also add sufficient
vector--like states charged under $U(1)'$ to give all the additional
matter (including states both in the $\tpt$ and $\fpf$'s, and the
$\Si$'s) masses. The derivation of the minimal set (in the sense of
reducing their contribution to the beta--function) of states and charges
that satisfies these conditions is a difficult problem
in general. As our interest is only in the value
of the minimal $U(1)'$ beta--function coefficient $b$ (including the
contributions from the SM-non-singlets states) we just quote
the results for $b_{\rm min}$ for the various models in
Table~\ref{tab:bminfn}, and where we have employed an ansatz
for the spectrum of anomaly cancelling states\footnote{We doubt that it is
possible for some of the SM singlets to be very light, which would have 
reduced significantly the $\beta$-function coefficients $b_{\rm min}$.
Constraints on this possibility come
predominantly from supernova cooling and to a lesser extent 
big--bang nucleosynthesis (BBN). 
If these SM singlets are massless, they will be 
produced copiously inside supernovae through their $Z'$ interactions.  
Once produced, they will free stream out of the supernova leading 
to rapid cooling.  Consistency with SN1987A observation requires that 
the $Z'$ mass must be greater than about $1\tev$ or that the singlet states
must be heavier than about $30\mev$.}.
(Our ansatz is to choose a set of $U(1)'$-charged states, $\Sigma$, 
which cancel the extra 
anomalies and simultaneously contribute minimally to the $U(1)'$ 
$\beta$-function. We then include a minimal set $U(1)'$ vector-like states 
which give mass to the $\Sigma$'s.)

\begin{table}
\centering
\begin{tabular}{|c|cccccc|}
\hline
Model&A&B($-1$)&B(7/5)&C($-1$)&C(7/5)&C(1) \\
\hline
$b_{\rm min}$ & 1363 & 280 & 174 & 129 & 154 & 191  \\
\hline
\end{tabular}
\caption{Minimal beta-function coefficients (in the
normalization $x=1$) for the models
defined in the text, together with additional SM--singlet
matter to cancel $[U(1)']^3$ and gravitational anomalies, and give mass to all
non-MSSM states. The version of Model A considered has a single
$\fpf$.}
\label{tab:bminfn}
\end{table}

Strictly speaking our ``unification principle'' does not absolutely
require the pertubativity of $U(1)'$ up to the GUT scale -- it is only
the SM gauge couplings that we require to successfully unify while
still perturbative. For instance, it is possible
that our extra $U(1)'$ gauge symmetry is enhanced into a non-Abelian
gauge symmetry well before the GUT scale, in which case the following
is (possibly much) too severe a restriction. Nevertheless it is 
interesting to see the bounds on the mass of the $Z'$ that follow
from such a requirement.

The restriction is derived as follows:
Using the Eqs.~(\ref{eq:xibar}) and (\ref{eq:delrhoM}) for the 
fitted quantities $\bar\xi$ and $\De\rho_M$, we find that for
the $x=1$ normalization choice,  
\beq
\al'(M_Z) \equiv {g'^2\over 4\pi}\simeq 4.43\times10^{-2}
{(\bar\xi)^2 \over \De\rho_M} \left({M_{Z'}\over M_Z}\right)^2.
\label{eq:alprime}
\eeq
However requiring that the Landau pole does not occur until a scale
$\La$ gives (at one loop) the restriction
\beq
\al'(M_Z) \leq {2\pi\over b} {1\over \log(\La/M_Z)},
\label{eq:landau}
\eeq
where $b$ is the beta--function coefficient.
Putting these two equations together leads to a restriction on the
$Z'$ to $Z$ mass ratio in terms of the ``measured'' quantities $\bar\xi$
and $\De\rho_M$, and the coefficient $b$ (for which we have a {\it lower}
bound given the minimal spectrum of $U(1)'$ charged particles necessary
for anomaly cancellation, \etc):
\beq
\left({M_{Z'}\over M_Z}\right)^2 \leq 142 {\De\rho_M \over (\bar\xi)^2}
{1\over b\log(\La/M_Z)}.
\label{eq:pertconstr}
\eeq
For the most restrictive case of $\La=2\times10^{16}~\gev$, this gives
\beq
\left({M_{Z'}\over M_Z}\right)^2 \leq 4.3 {\De\rho_M \over
(\bar\xi)^2 b_{\rm min}}.
\label{eq:pertGUT}
\eeq

\subsection{Experimental Constraints} \label{sec:expts}

Having defined each class of models, we know that each will, by definition,
be leptophobic. However it remains to be seen if they can describe the
physics as observed at LEP any better than the SM.
Note that as far as the agreement with the LEP data is concerned, the
only important feature of a model is the value of the ratio $y/x$.
(In all models except the $\eta$-model of Section~\ref{sec:etamodel}
we will choose to
normalize the $U(1)'$ gauge coupling $g'$ such that the quark
doublet charge $x=1$.)

To study this question,
we have performed a $\chi^2$ fit of each model to the LEP data, broadly
following the procedure of Refs.~\cite{altar,zmix}. We take 9 independent LEP
observables as inputs:
$\Gamma_Z$, $R_\ell=\Gamma_{\rm had}/\Gamma_\ell$, $\sigma_{\rm had}$,
$R_b$, $R_c$, $M_W/M_Z$, $A^b_{FB}$, $A^c_{FB}$, and $A^\ell_{FB}$. 
Theoretically, the
shift in each observable $\CO$ can be expressed as a function of
$\Delta\bar\rho$, $\bar\xi$, $x$, and $y$:
\beq
\frac{\Delta\CO}{\CO}=A_\CO \Delta\bar\rho + \left(B^{(1)}_\CO x
+B^{(2)}_\CO y\right)\bar\xi.
\label{eq:fit1}
\eeq
However, it is only in the simple case of no kinetic--mixing that
expressions for $A_\CO$ and $B^{(i)}_\CO$
follow directly from those given in Refs.~\cite{altar,zmix}. 
This is because they take Eq.~(\ref{eq:swsqb}) as the relation between 
$\swsqb$ and $\Delta\rho_M$; that is, the expressions of
Refs.~\cite{altar,zmix} assume that $\delta=0$.
For $\delta\neq0$, Eq.~(\ref{eq:swsqb2}) holds
instead. We then re-express 
\beq
A_\CO \Delta\bar\rho = A^{(1)}_\CO \Delta\bar\rho + A^{(2)}_\CO \Delta\swsqb,
\quad\quad
\Delta\swsqb\equiv\swsqb-\swsqb|_{\xi=\delta=0}
\label{eq:fit2}
\eeq
where $A^{(1)}_\CO$ includes only the {\sl explicit} dependence of the 
observable $\CO$ on $\Delta\bar\rho$, not the implicit dependence through
$\Delta\swsqb$. The coefficients $A^{(i)}_\CO$ are easily generalized from
the discussion of Ref.~\cite{zmix}; numerical values for the $A^{(i)}_\CO$
and $B^{(i)}_\CO$ are given in Table~\ref{table:coeffs}.
Note that $\Delta\swsqb$ is not a new parameter to be fit, since
it is simply a function of $\Delta\rho_M$, $\xi$ and $\delta$ 
through Eqs.~(\ref{eq:drho2}) and (\ref{eq:swsqb2}).
Clearly for $\delta=0$ the procedure here reduces to that of 
Refs.~\cite{altar,zmix}.
\begin{table}
\centering
\begin{tabular}{|c|rrrr|}
\hline
$\CO$ & $A_\CO^{(1)}$ & $A_\CO^{(2)}$ & $B_\CO^{(1)}$ & $B_\CO^{(2)}$ \\ 
\hline
$\Ga_Z$ & 0.98 & $-1.02$ & $-0.55$ & 0.50 \\
$R_\ell$ & $-0.04$ & $-0.83$ & $-0.78$ & 0.71 \\
$\sigma_{\rm had}$ & 0.006 & 0.12 & 0.32 & $-0.29$ \\
$R_b$ & 0.007 & 0.16 & $-2.8$ & $-0.71$ \\
$R_c$ & $-0.004$ & 0.33 & 5.4 & 1.4 \\
$M_W/M_Z$ & 0.38 & $-1.0$ & 0 & 0 \\
$A^b_{FB}$ & 0 & $-56$ & $-2.1$ & 0 \\
$A^c_{FB}$ & 0 & $-59$ & 2.4 & $-5.4$ \\
$A^\ell_{FB}$ & 0 & $-115$ & 0 & 0 \\
\hline
\end{tabular}
\caption{Coefficients $A_\CO$ and $B_\CO$ and observables $\CO$ used in the
fit to the electroweak data, as defined in Eqs.~(\protect\ref{eq:fit1})--
(\protect\ref{eq:fit2}).}
\label{table:coeffs}
\end{table}

Unlike Ref.~\cite{altar}, we have opted against using the data from SLC.
As is well known, the SLC data is approximately $2\sigma$ from the 
corresponding data at LEP. This could be a systematic effect at LEP,
SLC (or both), or a sign of new physics. Here we will take this
discrepancy not to be a sign of new physics. Therefore, as the
effects we are studying ($R_b$ and $R_c$) are in the LEP data, 
we choose, in this paper, to exclude the SLC data from our fits.

In our fits for the models of this Section, we have taken $S_{\rm extra}=0$ 
and allowed
for $\De\rho_{\rm extra}$ to be either zero or $-0.001$ consistent with our
discussion in Section~\ref{sec:oblique}. The negative value of 
$\De\rho_{\rm extra}$
in particular leads to a relaxation of the mass limits on the $Z'$.

\begin{figure}
\centering
\epsfxsize=3.5in
\hspace*{0in}
\epsffile{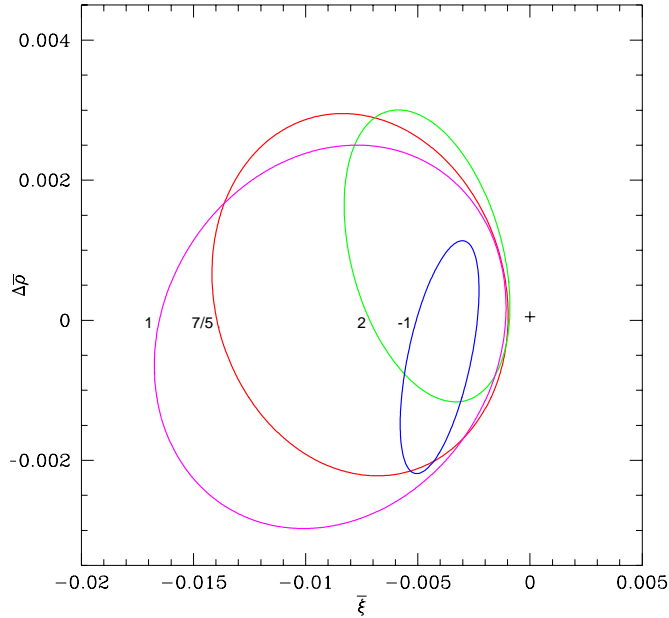}
\caption{99\% C.L. contours for the four basic classes of models labeled
by their $Q/u^c$ charge ratio in the 
$(\bar\xi,\Delta\bar\rho)$ plane. The cross represents the SM.}
\label{fig:all}
\end{figure}

In Table~\ref{table:fits}\ we have shown the $\chi^2$ for each of
the possible charge ratios $y/x=2,-1,7/5$, and $+1$ in addition to
the SM; the SM is defined by setting $\bar\xi=0$ in the fit. 
\begin{table}
\centering
\begin{tabular}{|c|cccc|} \hline
Model & $\Delta\bar\rho$ & {\footnotesize $\bar\xi$}
& $\chi^2$ & $\al_s(M_Z)$\\
\hline
SM & $5 \times 10^{-5}$ & 0 & 22.8 & 0.125 \\
2 & $9.1\times10^{-4}$ & $-4.6\times10^{-3}$ & 10.9 & 0.125\\
$-1$ &$-5.6\times10^{-4}$& $-4.1\times10^{-3}$ &14.8&0.110\\
7/5 & $3.5\times10^{-4}$ & $-7.6\times10^{-3}$ & 5.4 & 0.125\\
$+1$ & $-2.6\times10^{-4}$ & $-8.9\times10^{-3}$ & 4.0 & 0.123 \\ \hline
\end{tabular}
\caption{Results of fit to LEP data in the Standard Model (at
$\al_s=0.125$, the best fit for the LEP data alone) and
models with charge ratios $y/x=2,-1,7/5,+1$.
In all cases the $\chi^2$ are for 7 dof,
and $m_t=175$ \gev\ and $m_{\rm Higgs}=120$ \gev\
are assumed. The best fit value of $\al_s$ in the range $0.110$
to $0.125$ is quoted in each case.}
\label{table:fits}
\end{table}
For each model, we have given the values of $\De\bar\rho$ and $\bar\xi$
at the minimum $\chi^2$, as well as the value of $\al_s$ in the range
$0.110\le\al_s\le0.125$ which produces the best fit to the data.
For two of the models listed, the best fit value of $\De\bar\rho$ is negative;
however, the fit depends only weakly on $\De\bar\rho$ so that positive values
of $\De\bar\rho$ are allowed at relatively low $\chi^2$ as shown in 
Figure~\ref{fig:all}.

\begin{figure}
\centering
\epsfxsize=3.5in
\hspace*{0in}
\epsffile{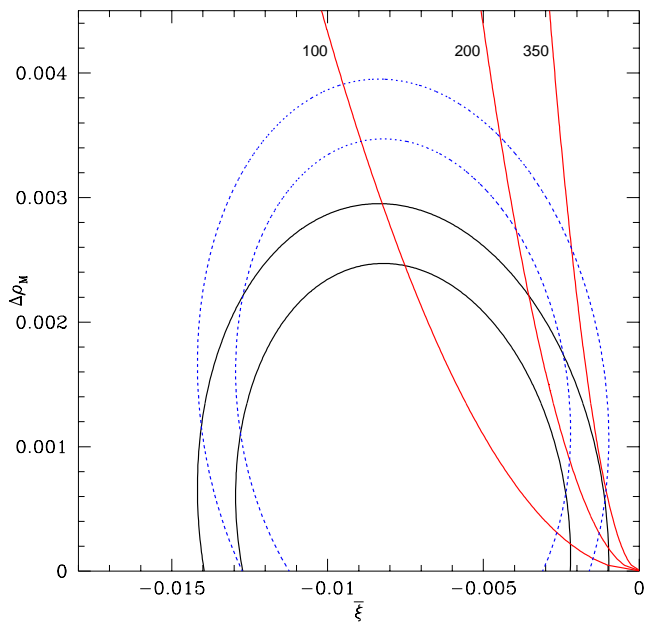}
\caption{$\chi^2$ contours for the C(7/5) Model in the
$(\bar\xi,\Delta\rho_M)$ plane. The solid ellipses
represent the 95\% and 99\% C.L. bounds on the fit. The dashed
ellipses represent the corresponding bounds if $\Delta\rho_{\rm extra}
=-0.001$. The three solid lines are contours of $M_{Z'}$ arising from the
theoretical constraint of perturbativity of the $U(1)'$ coupling
up to the GUT scale, and are labeled in GeV.}
\label{fig:C75}
\end{figure}

\begin{figure}
\centering
\epsfxsize=3.5in
\hspace*{0in}
\epsffile{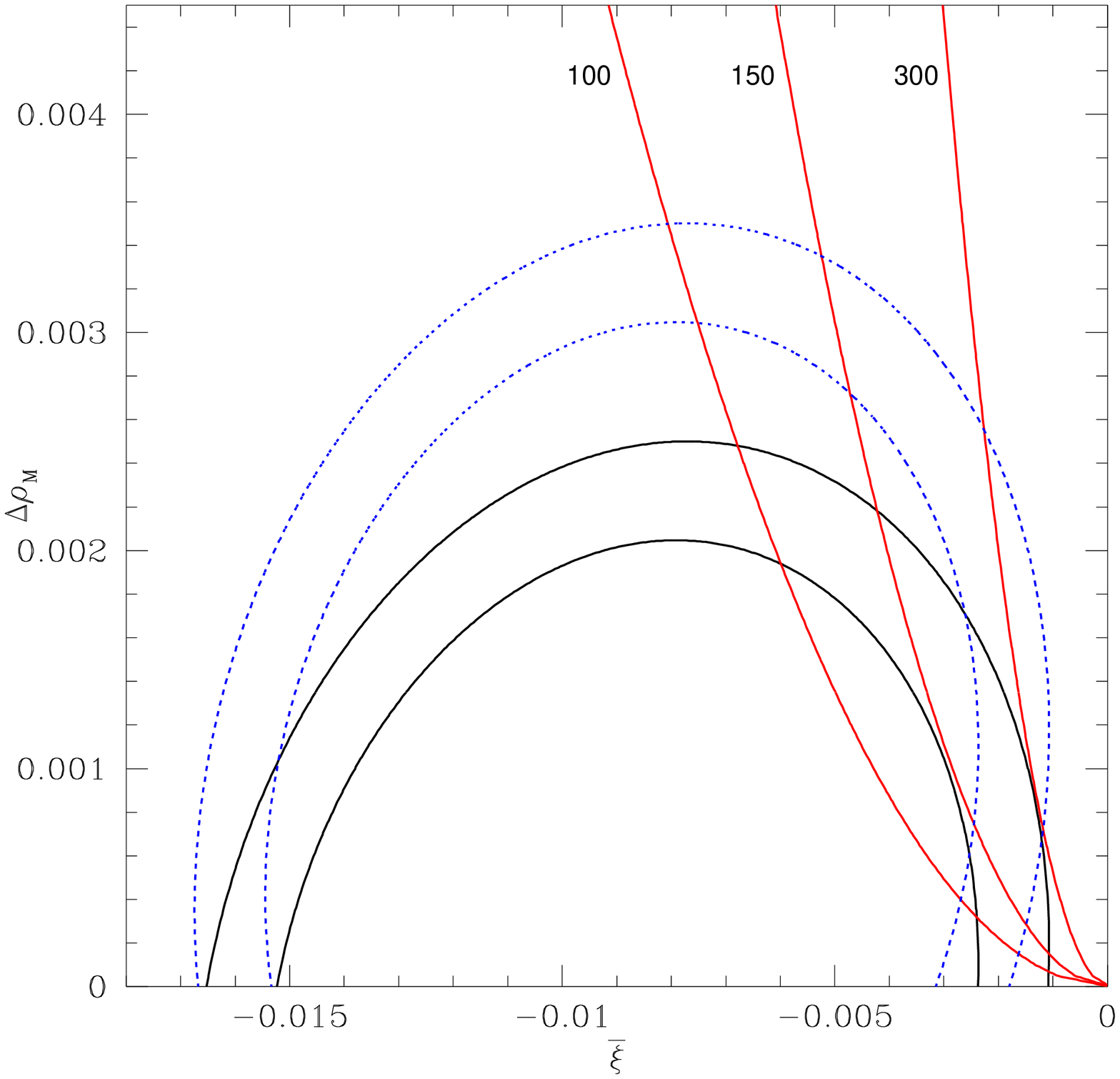}
\caption{$\chi^2$ contours for the C(1) Model 
in the $(\bar\xi,\Delta\rho_M)$ plane. 
See caption of Figure~\protect\ref{fig:C75}\ for explanation.}
\label{fig:C1}
\end{figure}

For the two most attractive models, C(7/5) and C(1), we have included plots
in Figures~\ref{fig:C75}\ and~\ref{fig:C1}\ of iso-$\chi^2$ contours
in the $(\bar\xi,\De\rho_M)$ plane. The solid ellipses represent contours
of $\chi^2=14.1$ and $18.5$, values which correspond to goodness-of-fits
of 95\% and 99\% respectively for 7 dof, assuming $\De\rho_{\rm extra}=0$.
In both cases, the contours impinge significantly into the physical
$\De\rho_M>0$ region. The dashed ellipses represent the case for which
$\De\rho_{\rm extra}=-0.001$ as discussed earlier in the text; for this case
the allowed values of $\De\rho_M$ are larger.

Figures~\ref{fig:C75}\ and~\ref{fig:C1}\ also show contours of constant
$M_{Z'}$ calculated assuming the perturbativity constraints of 
Eq.~(\ref{eq:pertGUT}) and using the values of $b_{\rm min}$ tabulated in
Table~\ref{tab:bminfn}. For the C(7/5) model, the 95\% (99\%) C.L.
bound on $M_{Z'}$ is $180\,(350)\gev$ for $\De\rho_{\rm extra}=0$ and
$250\,(500)\gev$ for $\De\rho_{\rm extra}=-0.001$. Similarly, for the
C(1) model the 95\% (99\%) C.L.
bound on $M_{Z'}$ is $150\,(300)\gev$ for $\De\rho_{\rm extra}=0$ and
$220\,(450)\gev$ for $\De\rho_{\rm extra}=-0.001$. The B(7/5) model
has mass limits only slightly stronger than those of the C(7/5) model:
$170\,(320)\gev$ for $\De\rho_{\rm extra}=0$. For the remaining models
in Table~\ref{tab:bminfn}, the corresponding $Z'$ mass limits are much 
stronger (with the exception of the $\eta$-model of Section~\ref{sec:etamodel},
which falls into the broad class of model A but has smaller value for the
$\beta$-function coefficient $b$). 

One might expect that $Z'$-models of the type considered here
would be strongly constrained by either UA2 or CDF/DO. However, the strongest
$Z'$ mass bounds in the literature depend on observation of the leptonic 
decays of the $Z'$, which are highly suppressed in these leptophobic models.
The dijet decays of the $Z'$, which dominate its width, are hard to detect 
above background except for limited ranges of $Z'$ masses and couplings.
In particular, CDF can only exclude $Z'\to jj$ for $M_{Z'}$ roughly
between 400 and $460\gev$~\cite{CDFzprime}, and then only for SM strength
(or stronger) couplings. UA2 has a similar bound of 
$M_{Z'}>260\gev$~\cite{UA2}, but here again one requires SM strength couplings.
Note that because of the small couplings that result from our
perturbativity constraint, we tend to find that the production cross-section
for the $Z'$ at a hadron collider is suppressed by at least 40\% compared
to the SM $Z$ cross-section.
We therefore find that UA2 does not provide a strong constraint on the
$Z'$ mass in these models.

All of the theoretical mass bounds that we have derived 
depend strongly on the value of the $U(1)'$ gauge coupling, and thus on
the size of $b_{\rm min}$ and especially on the assumption of 
perturbativity of the
$U(1)'$ gauge coupling all the way up to the GUT scale. If the $U(1)'$
interaction is enhanced to a non-Abelian group at some intermediate scale,
then the $Z'$ mass bounds are much weaker; we are investigating this
possibility. By either decreasing $b_{\rm min}$ or decreasing $\Lambda$
(the scale up to which we require perturbativity), $g'(M_Z)$ will increase.
As $g'$ increases the $Z'$ mass bound increases but the $Z'$ production
cross-section at a hadron collider, relative to a $Z$ of the same mass,
also increases. At some mass, however, the kinematic suppression of the
$Z'$ production wins and the experimental bound goes away. We will not
consider the details of these competing effects here.

Taking all the phenomenology together, including the possibility of naturally
small $Z$-$Z'$ mixing, we view the C(1), C(7/5), and the $\eta$-model of the
next Section as promising $Z'$ explanations of the $R_b$, $R_c$ anomalies.


\section{The $\eta$-model}\label{sec:etamodel}

As we noted in Section~\ref{sec:intro}, $E_6$ is a natural, and for
our purposes, minimal, choice for a simple GUT group containing
extra $U(1)$'s. In addition $E_6$ appears as an underlying feature
in many geometric compactifications of the $E_8\times E_8$ heterotic
string.  In either
case, the list of possible subgroups into which the $E_6$ can break is
small and well-defined.

Since $E_6$ is rank-6, its Cartan subalgebra contains two $U(1)$ generators
besides those of the SM gauge groups. At scales just above the electroweak
scale, the additional gauge symmetry could appear either as a commuting
$U(1)'$ factor (as we have been assuming up to this point) or as a unification
of the SM groups into some non-Abelian group (\eg, $SU(4)_c\times SU(2)_L
\times SU(2)_R$). The latter choice cannot describe the physics at LEP
since it cannot be leptophobic. Returning to the former, we can write
the new $U(1)'$ as a combination of the two extra $U(1)$'s in $E_6$,
usually denoted as $U(1)_\chi$ and $U(1)_\psi$:
\beq
Q'(\alpha)=\cos\alpha\,Q_\chi+\sin\alpha\,Q_\psi.
\eeq
In Table~\ref{table:E6}\ the charges $Q_\chi$ and $Q_\psi$ are given for
each of the states of the MSSM using the standard embedding into the
${\bf27}$. 
\begin{table}
\centering
\begin{tabular}{|c|ccc|c|}
\hline
 & $\sqrt{\frac{5}{3}}\,Y$ & $2\sqrt{6}\,Q_\psi$ & $2\sqrt{10}\,Q_\chi$ &
$2\sqrt{15}\,Q_\eta$ \\
\hline
$Q$ & $1/6$ & 1 & $-1$ & $-2$ \\
$u^c$ & $-2/3$ & 1 & $-1$ & $-2$ \\
$d^c$ & $1/3$ & 1 & 3 & 1 \\
$L$ & $-1/2$ & 1 & 3 & 1 \\
$e^c$ & 1 & 1 & $-1$ & $-2$ \\
$H_u$ & $1/2$ & $-2$ & 2 & 4 \\
$H_d$ & $-1/2$ & $-2$ & $-2$ & 1 \\
$D$   & $-1/3$ & $-2$ & 2 & 4 \\
$D^c$ & 1/3 & $-2$ & $-2$ & 1 \\
$\nu^c$ & 0 & 1 & $-5$ & $-5$ \\
$S$ & 0 & 4 & 0 & $-5$ \\
\hline
\end{tabular}
\caption{$U(1)$ charges of the states of a ${\bf 27}$ of $E_6$.}
\label{table:E6}
\end{table}

No linear combination of $U(1)_\chi$ and $U(1)_\psi$ is completely
leptophobic. The best one can do is to find models for which the axial
coupling of the charged leptons is zero. Since the vectorial contributions
for charged leptons appear proportional to $1-4\swsq\simeq0.07$, the
$Z'$ coupling to charged leptons could be highly suppressed with respect
to the hadronic couplings. However, such models would necessarily have
couplings to the neutrinos of order the hadronic couplings. If, after
$Z$-$Z'$ mixing the net effect were an increase in $\Gamma_{\rm inv}$ at LEP,
the model could be quickly ruled out. On the other hand, if $\Gamma_{\rm
inv}$ were to decrease, one could imagine that some new source of
invisible $Z$-decays (\eg, neutralinos) could offset the difference.
We consider such a scenario to be fine tuned and do not consider
it here.

However, as was discussed in Section~\ref{sec:rge}, in an arbitrary
$U(1)_a\times U(1)_b$ model, there is one more free parameter, a mixing
parameter $g_{ab}$ for the two groups. In the case of the breaking of
some unified gauge group, $G_{\rm GUT}$, at some high scale into
$G_{\rm GUT}\to SU(3)_c\times SU(2)_L\times U(1)_Y\times U(1)'$, the
value of $g_{ab}$ will be zero {\sl at the high scale}. Nonetheless, 
through its RGE's, Eq.~(\ref{eq:rge}), 
$g_{ab}$ will be driven to non-zero values for generic
particle content. The effective coupling to the $Z'$ is then
$\Qeff=Q'(\alpha)+\delta\,Y$ where $\delta=g_{ab}/g'$.

\begin{figure}
\centering
\epsfxsize=3.5in
\hspace*{0in}
\epsffile{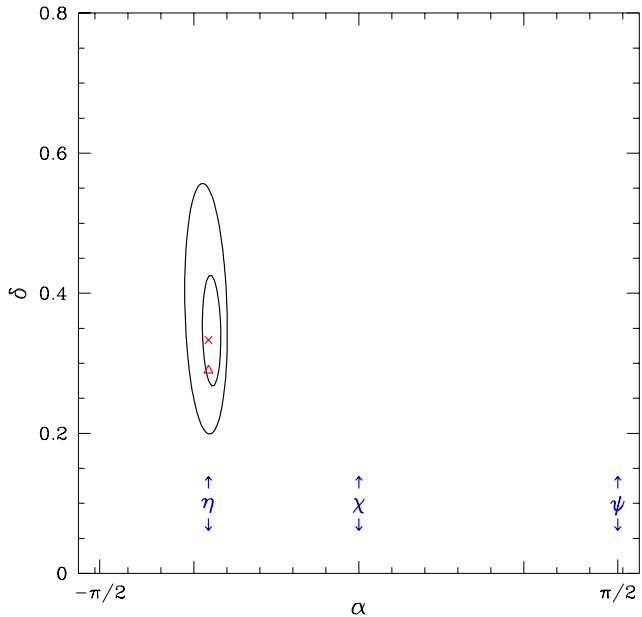}
\caption{$\chi^2$ contours for general $E_6$ models. The two contours represent
confidence levels of 95\% and 99\%. Three canonical $E_6$ models are labeled
at the bottom. The two points highlight the $\eta$-model with $\delta=1/3$
(${\bf\times}$) and $\delta=0.29$ (${\bf\triangle}$).}
\label{fig:E6}
\end{figure}

{}From the low-energy point of view, $\delta$ is a completely free parameter 
which must be fit to the data just as we did $\bar\xi$ or $\Delta\rho$.
Therefore, we have repeated the
$\chi^2$ analysis of the previous Section; however the charges
of the SM fermions are now completely determined in terms of $\alpha$ instead
of $x$ and $y$. Figure~\ref{fig:E6} is a $\chi^2$ plot in the plane of
$(\alpha,\delta)$ showing the fits to the LEP data at 95\%\ and 99\%\ C.L.\
At each point in the plane, the $\chi^2$ value is minimized with respect to
the remaining two free parameters, $\Delta\rho$ and $\bar\xi$.
Along the bottom of the plot are indicated the values of $\alpha$ consistent 
with the $\chi$, $\psi$, and $\eta$ models ($\alpha=0$, $\pi/2$,
$-\tan^{-1}\sqrt{5/3}\simeq-0.91$ respectively) commonly
discussed in the literature. All previous discussions of these models
(with the exception of Ref.~\cite{cvetic}) have tacitly taken $\delta=0$.

What is remarkable about the fit is that it picks a very
particular model out, for a limited range of $\delta$. To fall within the
95\%\ C.L. region ($\chi^2\le14.1$), 
a model must have $\alpha=-0.89\pm0.06$ and $\delta=0.35\pm0.08$.
Recall that the SM has a $\chi^2=22.8$ in the same parameterization.
Only one model lies within the region of allowed $\alpha$: the so-called 
$\eta$-model. The charges of the MSSM states under $U(1)_\eta$ 
are given in Table~\ref{table:E6}.

That the best fit in the ($\alpha,\delta$) plane lies at $Q'\simeq Q_\eta$
and $\delta\simeq1/3$ is not surprising. The effective charge $\Qeff
=Q_\eta+Y/3$ is completely leptophobic; in fact it is the only combination
of the three Abelian generators in $E_6$ which is leptophobic\footnote{After
submission of this paper, we were kindly informed by F.~del~Aguila that the
possibility of a leptophobic $U(1)$ in $E_6$ had been observed in
Ref.~\cite{delag2}; however, it was not realized that the required
value of $\delta$ was naturally generated through radiative effects in a model
with realistic matter content.}.
Note that the $Q_{\eta}$ charges of the lepton doublet $L$ and the lepton
singlet $e^c$ are proportional to their hypercharges.  Thus,
$U(1)_\eta$ is uniquely picked out as capable of describing the new physics
at LEP. In Figure~\ref{fig:E6}\ we have shown the $\delta=1/3$ $\eta$-model
with a cross.

If $U(1)'$ is indeed $U(1)_\eta$, there are a number of direct consequences
both for theory and phenomenology. First, $U(1)_\eta$ does not fit into any
GUT group smaller than $E_6$. Thus, if the unification of the gauge couplings
at a scale near $10^{16}\gev$ is not an accident, it indicates either a
true field-theoretic $E_6$ GUT (and {\sl no} $SU(5)$ or $SO(10)$
unification) or string-type unification in which $SU(3)\times SU(2)_L \times
U(1)_Y \times U(1)_\eta$ unifies directly at the scale $M_{\rm MSSM} 
=2 \times 10^{16}~\gev$. Second, cancellation of the anomalies in 
Eqs.~(\ref{eq:31})--(\ref{eq:1Y}) requires the existence of three
complete {\bf 27}'s of $E_6$.  
Besides the usual states of the MSSM, one can expect three pairs
of $D$ and $D^c$  quarks which are
$SU(2)_L$ singlets with $Y=\mp1/3$, two additional pairs of $SU(2)_L$ doublets
with $Y=\pm1/2$, three right-handed neutrinos, $\nu^c_i$, plus SM singlets
(at least one of which will receive a vev to break $U(1)_\eta$ and will
be eaten by the $Z'$).

We can now write the mass matrix of the $Z$-$Z'$ system. Defining $\tan\beta
\equiv\vev{H_u}/\vev{H_d}$ and $g_\eta$ to be $E_6$-normalized, 
the off-diagonal element in the mass matrix is 
given as in Eq.~(\ref{eq:mzz}):
\bea
\Delta m^2&=&\frac{2g_2g'}{\cw}\sum_i\vev{T_{3i}\phi_i}\vev{(Q_\eta+
\delta\,Y)\phi_i} \nonumber \\
&=&-\frac{1}{2\cw}\sqrt{\frac{5}{3}}g_\eta g_2 v_Z^2\sin^2\beta
\eea
where the last equality holds for the case where the only $SU(2)_L$ 
doublets with
non-zero vev's are $H_u$ and $H_d$. For the completely leptophobic 
$\eta$-model (\ie, $\delta=1/3$),
$\bar\xi$ and $\Delta\rho_M$ are then simply
\beq
\bar\xi=\frac{g^2_\eta\cwsq}{g_2^2}\sqrt{\frac{5}{3}}\sin^2\beta\left(
\frac{M_Z^2}{M_{Z'}^2}\right),
\quad\quad
\Delta\rho_M=\sqrt{\frac{5}{3}}\sin^2\beta\,\bar\xi\left(1-\frac{1}
{15\sin^4\beta}\right).
\label{eq:rhoxieta}
\eeq
Unfortunately, such a relationship between $\Delta\rho_M$ and $\bar\xi$
does not provide a very good fit to the data except near the unphysical
value of $\tan\beta\simeq0.6$;
the best fit consistent with Eq.~(\ref{eq:rhoxieta}) and $\tan\beta>1$
has $\chi^2$ of 22.0,
not much better than the SM $\chi^2=22.8$. There
is a second related problem: since $\Delta m^2\sim M_Z^2$ and we expect
(in the absence of tuning) for the $Z'$ mass to be only somewhat heavier,
we should expect large mixing angles $\xi$ to result. This is generic problem
of $U(1)'$ models where the $U(1)'$ is expected to be radiatively broken 
close to the weak scale~\cite{cvetic2}.

The solution to both problems involves the introduction of additional
$SU(2)_L$ doublets, charged under $U(1)'$, which receive vev's near the
weak scale. In our case these will play several roles: arranging the
$\beta$-functions of the model to unify at the GUT scale, allowing for
small $\xi$ by cancelling the $H_u$ contribution to $\Delta m^2$, 
likewise decoupling $\Delta\rho_M$ from $\bar\xi$, and driving $\delta>0$.

Consider, for example, extending the minimal $\eta$-model to include 
the pair of doublets which fit into the $[{\bf78},{\bf16
+\bar{16}},\fpf]$ of $[E_6,SO(10),SU(5)]$, with the doublet in the ${\bf5}$
getting a vev, $v_\ell$, near the weak scale. Then in the leptophobic 
$\eta$-model, $\Delta m^2\propto (v_Z^2\sin^2\beta-v_\ell^2)$. If a near
cancellation can be arranged between the two terms in $\Delta m^2$, then
small mixing will result and simultaneously $\Delta\rho_M\ll\bar\xi$ as
needed phenomenologically. Since $M_Z^2\propto(v_Z^2+v_\ell^2)$ and we need
$v_Z$ and $v_\ell$ of the same order, the Higgs vevs, $v_u$ and $v_d$, 
which give masses to the fermions will be proportionally smaller. In the case
$v_d\ll v_u\sim v_\ell$, the large top-bottom mass ratio is natural and the
top Yukawa is of the same size as one would expect in the MSSM with 
$\tan\beta=1$. This is actually still below the top Yukawa infrared 
pseudo-fixed
point, which now takes a larger value ($h_t^{\rm fixed} \simeq 1.25$)
because of the slow running of $\alpha_s$ in this model.

Imposing on the superpotential of the minimal $\eta$-model a discrete 
$Z_2$ symmetry (a simple extension of the usual $R$-parity) one finds:
\beq
W_\eta=Qu^cH_u + Qd^cH_d + Le^cH_d + SH_uH_d + SDD^c + L\nu^c H_u
\eeq
Under the $R$-parity, all the states of the ${\bf 27}$ are odd except
$H_u$, $H_d$ and $S$. This superpotential forbids dimension-4 proton decay;
dimension-5 operators are also known to be unobservably small in the
$\eta$-model~\cite{pdecay}. There appears in the superpotential a
Yukawa mass term for the right-handed neutrino fields, $L\nu^cH_u$.
To be consistent with current neutrino mass bounds, this coupling must be small
or zero or the $\nu^c$ must have large Majorana mass terms through some 
singlets.   By flipping the $R$-parity assignment of the $\nu^c$ one can
forbid the term altogether, but at the price of introducing into the
superpotential the term $\nu^cDd^c$. Such a term would lead to 
$D$-$d^c$ mixing were $\nu^c$ to receive a non-zero vev. 

One can also expect radiative symmetry breaking
much as in the MSSM. If the $SDD^c$ coupling is $\CO(1)$, the soft mass
term for the $S$-field, $m_S^2$, will be driven negative through its RGE's,
triggering $U(1)_\eta$-breaking through $\vev{S}\neq0$ 
at a scale just above the electroweak scale.
(The electroweak symmetry will similarly be broken by $m^2_{H_u}$ running
negative due to the large top Yukawa coupling.)   Since the singlet $S$ has
no electroweak interactions unlike $H_u$, it is conceivable that the 
mass-squared of the $S$ fields turns negative at a larger momentum scale 
compared to $H_u$.  
The non-zero $\vev{S}$ will
in turn produce a $\mu H_uH_d$ and a $\mu'DD^c$ term. For 
$SH_uH_d$ and $SDD^c$ couplings of $\CO(1)$, one expects $\mu,\mu'\sim
M_{Z'}$. In particular, it is natural for the $D$ and $D^c$ states to be
heavier than the $Z$. Finally, we note that there is no mechanism within the 
$\eta$-model for $\nu^c$ to receive a vev radiatively which does not violate
some other constraint (such as neutrino mass bounds)~\cite{pdecay}. Thus 
$D$-$d^c$ mixing will not occur.

The $\eta$-model with only three ${\bf 27}$'s of $E_6$ does not
satisfy all of our initial principles because it does not have gauge coupling
unification. As mentioned above, unification can be arranged by introducing
one pair of $SU(2)_L$ doublets with hypercharges $\sqrt{5/3}\,Q_Y=
\pm\frac{1}{2}$. From a string point of view, these may be viewed as coming
from a ${\bf\bar{27}}+{\bf27}$ or a ${\bf78}$, 
the rest of whose states received masses
at the string scale~\cite{witten2}. This, along with anomaly cancellation
considerations, requires the doublets to have equal and opposite $Q_\eta$.
If these doublets also have non-zero effective charges $Q_\eta+\delta\,Y$, 
their vev's may contribute to the $Z$-$Z'$ mixing matrix as outlined above.
A problem may potentially arise in trying to generate vev's for these doublets
radiatively; one possibility is to allow couplings of the type $H_u H_d'$
through singlets.  

(This model has, beyond the spectrum of the MSSM,
three each of $({\bf3,1})$ and $({\bf\bar3,1})$ and six of $({\bf1,2})$.
This is exactly the content of three $(\fpf)$'s of $SU(5)$. 
Note that in terms of the charge ratio $y/x$, the purely leptophobic
($\delta=1/3$) $\eta$-model is equivalent to Model A of 
Section~\ref{sec:models}. However, the presence of kinetic mixing 
($\delta\neq0$) induces contributions to the oblique electroweak parameters
not present in Model A.
Also unlike the purely leptophobic models of that Section the value of
$\delta$ in the $\eta$-model is generically not $1/3$, but is instead 
determined through the RGE's and thus through the low-energy spectrum.
Further, its $\beta$-function is substantially smaller than that of Model
A with a single $(\fpf)$, since for the $\eta$-model the anomaly cancellation
is generation by generation, providing a more economical set of charges.)

There are two variants of the $\eta$-model for which the value of $\de$ at
the electroweak scale is of particular interest: 
{\sl (i)}~The ``minimal'' $\eta$-model
that possesses three generations of ${\bf 27}$'s and one additional
vector-like pair of Higgs doublets that arises from the ${\bf 78}$ of
$E_6$. These doublets have charges $\sqrt{5/3}\,Q_Y=-1/2$
and $2\sqrt{15}\,Q_\eta=6$ under the GUT-normalized $U(1)_Y\times U(1)_\eta$
symmetries; {\sl (ii)}~The ``maximal'' $\eta$-model with in addition to the
states of the minimal $\eta$-model a further effective
$\fpf$ of $SU(5)$ is added (so that unification is preserved), but which is
composed of a second vector-like pair of the doublets in the ${\bf 78}$
together with the color triplets $D+\bar D$ coming from the ${\bf 27}+
{\bf\bar{27}}$. The maximal model has the largest field content
consistent with perturbative unification of the gauge couplings at
$2\times 10^{16} \gev$. The values of the charge inner products $B_{ij}$
for these two models are given in Table~\ref{table:betafns}.
The field content of both these models is consistent with small $\Delta m^2$
in the $Z$-$Z'$ mass matrix.

\begin{table}
\centering
\begin{tabular}{|c|ccc|}
\hline
Model & $B_{YY}$ & $B_{\eta\eta}$ & $B_{Y\eta}$ \\
\hline
$\eta_{\rm min}$ & $9+\frac{3}{5}$ & $9+\frac{12}{5}$ & $-\frac{6}{5}$  \\
\hline
$\eta_{\rm max}$ & $9+\frac{8}{5}$ & $9+\frac{32}{5}$ & $-\frac{16}{5}$ \\
\hline
\end{tabular}
\caption{Beta-function coefficients for the Minimal and Maximal $\eta$-models,
GUT normalized.}
\label{table:betafns}
\end{table}

Running the SM couplings up to the unification point and then numerically
running
the RGE's of Eq.~(\ref{eq:rge}) for $g_Y$, $g_\eta$ and $g_{Y\eta}$ down to 
the electroweak scale, we find predictions for $\de$ in the two models:
\beq
\de_{\rm min} = 0.11, \quad\quad\quad \de_{\rm max} = 0.29.
\label{eq:delres}
\eeq
Both of these are calculated with $\al_s(M_Z)=0.120$. Larger values
of $\al_s(M_Z)$ lead to a slight increase in the values of $\de$ 
compared to Eq.~(\ref{eq:delres}). The threshold corrections
to $\delta$ coming from mass splitting of the light states are typically
of order $0.01$.
It is quite remarkable that the totally leptophobic value
of $\de=1/3$ is very nearly predicted by the renormalization group
running of the ``maximal'' $\eta$-model. 
{}From the one-loop RGE's, the value of the 
$U(1)_\eta$ gauge coupling at the electroweak scale is $g_\eta = 0.40$. 

Given these values of $\delta$ we can now investigate how well the $\eta$-model
variants can fit the LEP data. As discussed in Section~\ref{sec:oblique}
we will consider both the case of $S_{\rm extra}=0$ and 
$S_{\rm extra}=0.14$ per pair of 
higgsino/lepton-like doublets. We will take $\De\rho_{\rm extra}=0$.
The minimal model is
clearly disfavored by the data, having a $\chi^2$ no better than the SM for
both values of $S_{\rm extra}$. 
Likewise the maximal model with $S_{\rm extra}=0$ is disfavored.
The phenomenologically favored maximal model has 5 doublet pairs giving
$S_{\rm extra}=0.7$ and a minimum
$\chi^2=13.9$ at a $Z'$ mass of $215\gev$; this is
within the 95\% C.L. bounds shown in Figure~\ref{fig:E6}, where the model
is indicated by a triangle. At the minimum, $S\equiv S_M+S_{\rm extra}=-0.1$.
Note that the goodness of the fit does not depend strongly on the exact value
of $S_{\rm extra}$ in the range 0.5 to 1.5; in particular the resulting
$S$ only varies within the range $-0.1$ to $0.1$. 

\begin{figure}
\centering
\epsfxsize=3.5in
\hspace*{0in}
\epsffile{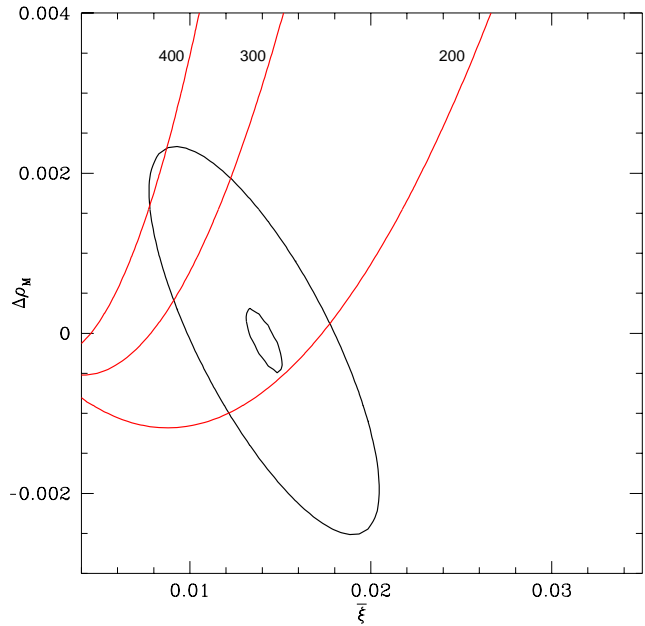}
\caption{$\chi^2$ contours for the $\eta$-model with $\delta=0.29$
in the $(\bar\xi,\Delta\rho_M)$ plane. 
See caption of Figure~\protect\ref{fig:C75}\ for explanation.
Additional positive contributions to $\Delta\bar\rho$ reduce the best
fit value of the $Z'$ mass.}
\label{fig:eta}
\end{figure}

Given $g_\eta$ and the bounds on $\Delta\rho_M$ and $\bar\xi$ we are in a 
position to calculate the bounds on the $Z'$ mass, using 
Eq.~(\ref{eq:delrhoM}). For the $\eta$-model with $\delta=0.29$, we find that
in order to fall within the 95\% (99\%) C.L. limits for our fit, then
$M_{Z'}\leq 240\,(420)\gev$, under the assumption of no additional 
contributions to $\Delta\bar\rho$. (New positive contributions to 
$\Delta\bar\rho$, which are natural in these models, push the best
fit $Z'$ mass to lower values.)
These fits are shown in Figure~\ref{fig:eta}. 
UA2 has performed a $Z'$ search in the dijet
channels, excluding a $Z'$ with 100\% branching fraction to hadrons and
SM strength interactions up to masses of $260\gev$~\cite{UA2}. 
However, given the
value of $g_\eta=0.4$ and the $U(1)_\eta$ charges of the quarks, one can
show that the production cross-section for this $Z'$ is approximately
1/4 that of the $Z$, too small to be excluded at UA2.

What is remarkable about this analysis is that the $\eta$-model, which has been
extensively studied in the literature and for which strong bounds on its mixing
with the $Z$ and its mass have been published, has been resuscitated by the
inclusion of the additional $U(1)$ kinetic mixing effect. 
This is even more so, 
since the value of $\delta$ is correctly predicted in specific models in which 
only one discrete choice of matter content has been made!

\section{Conclusions}

In this paper, we have investigated the possibility of explaining the $R_b$
excess -- $R_c$ deficit reported by the LEP experiments through $Z$-$Z'$ mixing
effects. We have constructed a set of models consistent with a restrictive set
of principles: unification of the SM gauge couplings, vector-like additional
matter, and couplings which are both generation-independent and leptophobic.
These models are anomaly-free, perturbative up to the GUT scale, and contain 
realistic mass spectra. Out of this class of models, we find three explicit
realizations (the $\eta$, C(7/5), and C(1) models)
which fit the LEP data to a far better extent than the
unmodified SM or MSSM and satisfy all other phenomenological constraints which
we have investigated. The $\eta$-model is particularly attractive, coming 
naturally from geometrical compactifications of heterotic string theory. 
This is especially so since the value of the mixing parameter, $\delta$, is
correctly predicted given only one discrete choice of matter content.

In general, these models predict extra matter below $1\tev$ and $Z'$ gauge
bosons below about $500\gev$, though the $Z'$ of these models will be difficult
to detect experimentally.

\section*{{\bf Note Added}}
After this work was completed two further interesting works concerning
the experimental consequences of leptophobic $U(1)$'s appeared
\cite{berkeley}\cite{barger}. These papers noted that there can exist
important low-energy constraints on leptophobic models arising from
atomic parity violation (APV) and deep-inelastic neutrino scattering
experiments. In particular, Ref.~\cite{berkeley} argued that the aesthetically
appealing models that we have constructed in this paper are strongly
disfavored by the APV data. While this is usually true in the heavy
$Z'$ mass approximation that we have been employing up to now, this
conclusion does not hold in the very interesting case of a light $Z'$
($M^2_{Z'}\gtap m^2_{Z}$), as we will now outline.

The APV experiments
result in constraints on the so-called weak nuclear charge $Q_W$ 
of various elements such as Cesium and Thallium with high atomic and neutron
numbers $Z$ and $N$. The charge $Q_W$
is itself defined in terms of the product of the axial electron coupling
with the up and down type quark vector coupling via
$Q_W = -2\left\{ C_{1u}(2Z+N) + C_{1d}(Z+2N)\right\}$ where
$$
\CL_{NC} = -{G_F\over\sqrt{2}}\sum_{i=u,d} C_{1i}
(\bar{e}\ga_\mu\ga_5 e)(\bar{q}_i \ga^\mu q_i)+\cdots
$$
In the case where the $M_{Z'}\sim m_Z$, both $Z_1$ and $Z_2$ exchange
contribute to the coefficients $C_{1i}$. In the approximation where
the mixing is small $\xi \ll 1$, but {\it no} expansion is made in the
mass ratio $r\equiv (m^2_{Z}/M^2_{Z'})$, the expression for the
$C_{1i}$'s is
$$
C_{1i} = -\left\{ v_i + v^{'}_i \bar{\xi} (1-r)\right\},
$$
where the $v$'s and $\bar{\xi}$ are defined in
Eqs.~(\ref{eq:newLnc}) and (\ref{eq:xibar}) respectively.
It is therefore clear that the constraint from the APV data
becomes vacuous as $r\to 1$. Specifically, we find
that the APV data do not significantly increase the total
$\chi^2$ for $Z'$ masses below about $150~\gev$. 

One may similarly consider the effect of a leptophobic
$Z'$ on the neutrino scattering experiments. We find that the parameters
$\ep^{i}_L$ and $\ep^{i}_R$ defined in Ref.~\cite{RPP}, are altered
by an amount
$$
\De\ep^{i}_{L/R} = \left({v^{'}_i \mp a^{'}_i\over 2}\right)\bar{\xi}(1-r),
$$
respectively. Thus the weaker constraints from the neutrino scattering
data also dissappear for light to moderate $Z'$ masses.

We will address the full fit including these constraints (as well
as the SLC and other data),
more fully in a forthcoming paper~Ref.~\cite{forthcoming}, where
we will also discuss the models with variant Higgs structure mentioned
in the Section~\ref{sec:models} footnote.

\section*{Acknowledgments}
We wish to thank K.~Dienes, S.~Martin, J.~Wells and F.~Wilczek for helpful
discussions, F.~del~Aguila for informing us of Ref.~\cite{delag2},
and especially B.~Holdom for important comments concerning
electroweak corrections that pertained to an earlier version of
this work.


\begin{thebibliography}{99}

\bibitem{lep}
	LEP Electroweak Working Group, report LEPEWWG/95-02 (August 1995).

\bibitem{Z0pole}
	All calculations in this paper were done using the Z0POLE program of
	B.~Kniehl and R.~Stuart, Comput. Phys. Commun. {\bf 72} (1992) 175.

\bibitem{CDFtop}
	F.~Abe, \etal (CDF Collaboration), \PRL{74}{95}{2626}.

\bibitem{SUSYRb}
	J.~Wells, C.~Kolda and G.~Kane, \PLB{338}{94}{219};\\
	D.~Garcia, R.~Jimenez and J.~Sola, \PLB{347}{95}{321};\\
	E.~Simmons and Y.~Su, report BUHEP-96-4 (February 1996)
	{\tt [hep-ph/9602267]}.

\bibitem{mixingRb}        
	B.~Holdom, \PLB{339}{94}{114}; \PLB{351}{95}{279};\\
        X. Zhang and B.L. Young, \PRD{51}{95}{6584};\\
        E. Ma and D. Ng, \PRD{53}{96}{255}; \\
        M. Carena, H.E. Haber and C. Wagner, report CERN-TH-95-311 
	(December 1995) {\tt [hep-ph/9512446]};\\
        T. Yoshikawa, report HUPD-9528 (December 1995) 
	{\tt [hep-ph/9512251]};\\
	P.~Bamert, C.~Burgess, J.~Cline, D.~London and E.~Nardi, report
	MCGILL 96-04 (February 1996) {\tt [hep-ph/9602438]}.


\bibitem{otherRbRc}
	E. Ma, UCRHEP-T-153 {\tt [hep-ph/9510289]}; \\
        G. Bhattacharya, G. Branco and W-S. Hou, report CERN-TH/95-326 
	(December 1995) {\tt [hep-ph/9512239]}; \\
        C.V. Chang, D. Chang and W-Y. Keung, report NHCU-HEP-96-1
        (January 1996) {\tt [hep-ph/9601326]}.

\bibitem{feng}
	J.~Feng, H.~Murayama and J.~Wells, report SLAC-PUB-95-7089 (January
	1996) {\tt [hep-ph/9601295]}.

\bibitem{altar}
	G.~Altarelli, N.~di~Bartolomeo, F.~Feruglio, R.~Gatto and M.~Mangano,
	 report CERN-TH-96-20 (January 1996) {\tt [hep-ph/9601324]}.

\bibitem{chia}
	P.~Chiappetta, J.~Layssac, F.~Renard and C.~Verzegnassi,
	report PM-96-05 (January 1996) {\tt [hep-ph/9601306]}.

\bibitem{stringuni}
	K.~Dienes and A.~Faraggi, \NPB{457}{95}{409};\\
	K.~Dienes, A.~Faraggi and J.~March-Russell, report IASSNS-HEP-95/25
	(October 1995) {\tt [hep-th/9510223]}.\\
	For a recent review of unification in the string framework see:
	K.~Dienes, report IASSNS-HEP-95/97 (February 1996)
	{\tt [hep-th/9602045]}.

\bibitem{witten}
	E.~Witten, report IASSNS-HEP-96/08 (February 1996) 
	{\tt [hep-th/9602070]}.

\bibitem{holdom}
	B.~Holdom, \PLB{166}{86}{196}.

\bibitem{delarge}
	F.~del~Aguila, G.~Coughlan and M.~Quiros, \NPB{307}{88}{633}.

\bibitem{E6review}
	For a review, see J.~Hewett and T.~Rizzo, \PRT{183}{89}{193}. 

\bibitem{zmix}
	G.~Altarelli, R.~Casalbuoni, D.~Dominici, F.~Feruglio
	and R,~Gatto, \MPL{A5}{90}{495}.

\bibitem{holdomrho}
	B.~Holdom, \PLB{259}{91}{329}.

\bibitem{delaguila}
        R.~Foot and X-G.~He, \PLB{267}{91}{509}; \\
	F.~del~Aguila, M.~Masip and M.~Perez-Victoria, report UG-FT-46-94
	(July 1995) {\tt [hep-ph/9507455]}.

\bibitem{drees}
	M.~Drees and K.~Hagiwara, \PRD{42}{90}{1709}.

\bibitem{haber}
	R.~Barbieri, M.~Frigeni, F.~Giuliani, and H.~Haber, 
	\NPB{341}{90}{309};\\
	G.~Altarelli, R.~Barbieri, and F.~Caravaglios, \PLB{314}{93}{357}.

\bibitem{maroy}
	E.~Ma and P.~Roy, \PRL{68}{92}{2879}.

\bibitem{CDFzprime}
	F.~Abe, \etal, (CDF Collaboration), \PRL{74}{95}{3539}.

\bibitem{UA2}
	J. ~Alitti, \etal, (UA2 Collaboration), \NPB{400}{93}{3}.

\bibitem{cvetic}
	F.~del~Aguila, M.~Cveti\v c and P.~Langacker, \PRD{52}{95}{37}.

\bibitem{cvetic2}
	M.~Cveti\v c and P.~Langacker, report IASSNS-HEP-95/90 (November
	1995) {\tt [hep-ph/9511378]}.

\bibitem{pdecay}
	B.~Campbell, J.~Ellis, K.~Enqvist, M.~Gaillard and D.~Nanopoulos,
	Int. J. Mod. Phys. {\bf A2} (1987) 831.

\bibitem{witten2}
	E.~Witten, \NPB{258}{85}{75}.

\bibitem{delag2} 
	F.~del~Aguila, M.~Quiros and F.~Zwirner, \NPB{287}{87}{419}.

\bibitem{berkeley}
	K.~Agashe, M.~Graesser, I.~Hinchliffe and M.~Suzuki,
	report LBL-38569 (April 1996), {\tt [hep-ph/9604266]}.

\bibitem{barger}
	V.~Barger, K.~Cheung and P.~Langacker, report MADPH-96-936
	(April 1996), {\tt [hep-ph/9604298]}.

\bibitem{RPP}
	L. Montanet, \etal, (Particle Data Group), \PRD{50}{94}{1173}, 
	Section~26.2.

\bibitem{forthcoming}
	K.~S.~Babu, C.~Kolda and J.~March-Russell, report IAS-HEP-96/45
	(in preparation).

\end{thebibliography}
\end{document}